\documentclass[floatfix,10pt,aps,prb,twocolumn,showpacs,a4paper,superscriptaddress]{revtex4-1}

\usepackage{amsmath}   
\usepackage{mathtools} 
\usepackage{amssymb}   
\usepackage{mathrsfs}  
\usepackage{textcomp}  
\usepackage{accents}   
\usepackage{bm}        
\usepackage{soul}
\usepackage{relsize}
\usepackage{graphicx}  
\usepackage{epstopdf}  
\usepackage{subfigure} 
\usepackage{makecell}
\usepackage{hhline}
\usepackage{cellspace}
\usepackage{array}     
\usepackage{tabularx}  
\usepackage{multirow}  
\usepackage{booktabs}  
\usepackage{dcolumn}   
\usepackage{gensymb}   
\usepackage{setspace}
\usepackage{scrextend}
\usepackage{graphicx}
\usepackage[usenames,dvipsnames]{color} 
\usepackage{url}       
\usepackage{hyperref}  
\hypersetup{colorlinks,citecolor=Violet,linkcolor=Red,urlcolor=Blue}

\graphicspath{{./}}

\usepackage{soul}
\setstcolor{red}
\setul{}{1.5pt}
\usepackage[normalem]{ulem}
\newcommand\redsout{\bgroup\markoverwith{\textcolor{red}{\rule[0.5ex]{2pt}{1.4pt}}}\ULon}

\newcommand{\ymno}{Y$_{2}$Mn$_{2}$O$_{7}$}
\newcommand{\ymoo}{Y$_{2}$Mo$_{2}$O$_{7}$}
\begin{document}
\setlength{\heavyrulewidth}{0.08em}
\setlength{\lightrulewidth}{0.05em}
\setlength{\cmidrulewidth}{0.03em}
\setlength{\belowrulesep}{0.65ex}
\setlength{\belowbottomsep}{0.00pt}
\setlength{\aboverulesep}{0.40ex}
\setlength{\abovetopsep}{0.00pt}
\setlength{\cmidrulesep}{\doublerulesep}
\setlength{\cmidrulekern}{0.50em}
\setlength{\defaultaddspace}{0.50em}
\setlength{\tabcolsep}{4pt}

\title{\textit{Ab initio} determination of magnetic ground state of pyrochlore Y$_{2}$Mn$_{2}$O$_{7}$ }

\author{Mohammad Amirabbasi}
\affiliation {Department of Physics, Isfahan University of Technology (IUT), Isfahan 84156-83111, Iran}

%
\author{Mojtaba Alaei}
\email {m.alaei@iut.ac.ir}
\affiliation{Department of Physics, Isfahan University of Technology (IUT), Isfahan 84156-83111, Iran}
%

%

\date{\today}
\begin{abstract}
There are two discrepant experimental results on the magnetic ground state of Y$_{2}$Mn$_{2}$O$_{7}$, 
one study proposes a spin glass state, while another introduces the material as a ferromagnet.
In this study, we attempt to resolve this issue by employing density functional theory and Monte Carlo simulations. 
We derive different spin models by varying the Hubbard $U$ parameter in {\it ab initio} GGA+$U$ calculations. 
For the most range of Hubbard $U$, We obtain that the leading terms in the spin Hamiltonian are bi-quadratic and the nearest neighbor Heisenberg exchange interactions.
By comparing Monte Carlo simulations of these models with the experiments,
we find a ferromagnetic ground state for Y$_{2}$Mn$_{2}$O$_{7}$ as the most compatible with experiments. 
We also consider Y$_{2}$Mo$_{2}$O$_{7}$ as a prototype of the defect-free pyrochlore system with spin-glass behavior 
and compare it with Y$_{2}$Mn$_{2}$O$_{7}$.
The orbital degrees of freedom are considered as a leading factor in converting a 
defect-free pyrochlore such as Y$_{2}$Mn$_{2}$O$_{7}$ to a spin glass system. 
By changing the $d$ orbital occupations of Mo atoms, our GGA+$U$ calculations for Y$_{2}$Mo$_{2}$O$_{7}$ 
indicate many nearly degenerate states with different $d$ orbital orientations which reveals $d$ orbital degrees of freedom in this material.
While for Y$_{2}$Mn$_{2}$O$_{7}$, we find a single ground state with a fixed orbital orientation. 
Consequently, all of our {\it ab initio} approaches confirm Y$_{2}$Mn$_{2}$O$_{7}$ as a ferromagnetic system.
\end{abstract}

\maketitle

\section{introduction}
\label{sec:introduction}
\begin{figure}[b]
  \centering
  \includegraphics[width=0.8\columnwidth]{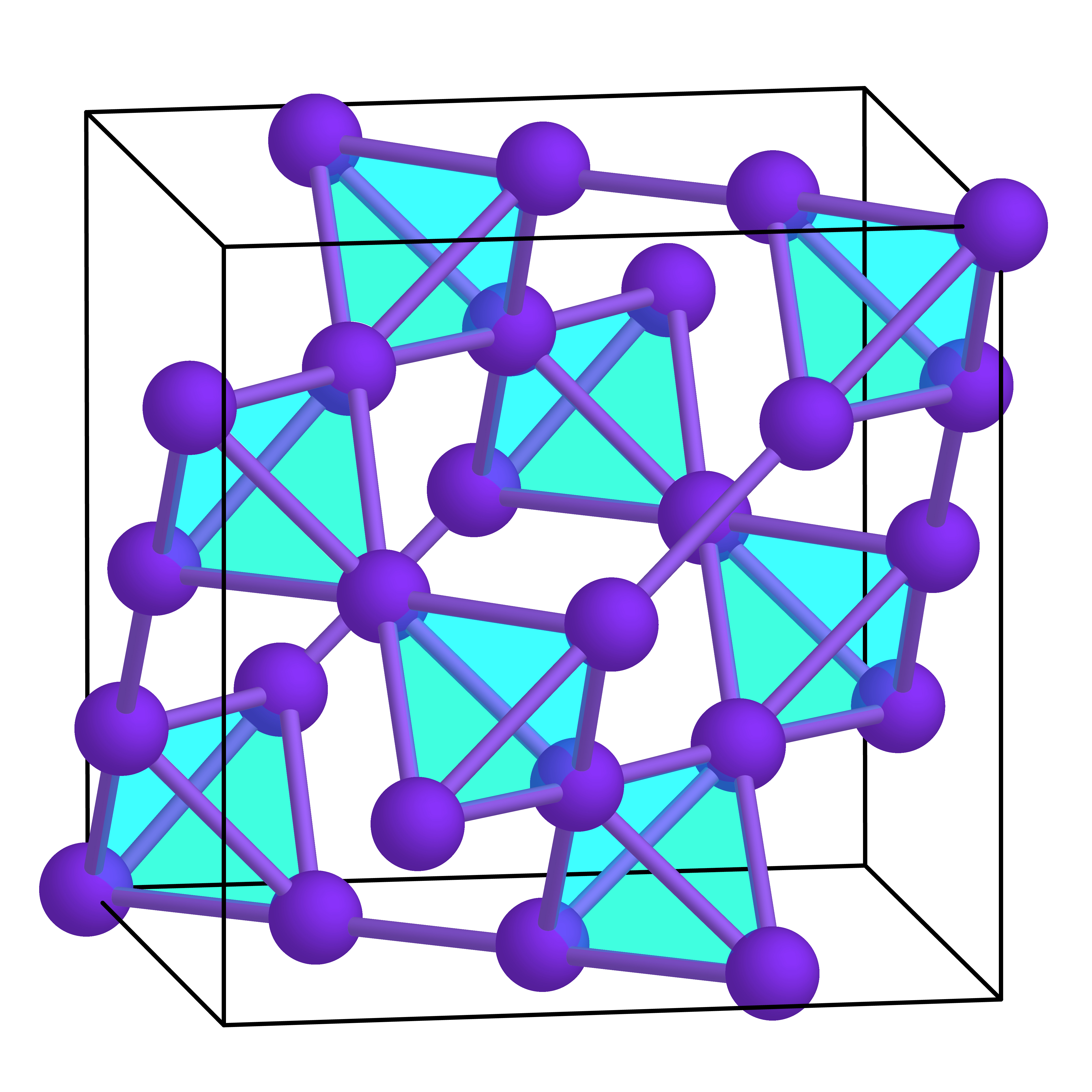} 
  \caption{(Color online) The Corner sharing tetrahedral network of Mn (Mo) atom in cubic pyrochlore
    Y$_{2}$Mn$_{2}$O$_{7}$ (Y$_{2}$Mo$_{2}$O$_{7}$).} 
  \label{fig:pyrochloe}
\end{figure}

The geometrical frustration~\cite{Lacroix} of magnetic pyrochlore oxides, with the chemical formula A$_{2}$B$_{2}$O$_{7}$~\cite{Greedan2001,Gardner2010}, 
is absorbing more researches on these materials. 
The geometrical frustration is due to the networks of corner-sharing tetrahedrons (Fig.~\ref{fig:pyrochloe} indicates one of these network), 
which are formed by the magnetic ions residing on A and B sites. 
Due to the diverse nature of the magnetic ions, their interactions, and geometrical frustration 
the ground state can be spin ice~\cite{Savary2017, Gaudet2019}, spin liquid~\cite{Balents2010}, or
spin glass~\cite{Davies2019}. 
If this frustration accompanies randomness such as chemical disorders  
and crystallographic defects, magnetic moments of ions will
freeze with random spatial orientations 
~\cite{Singh2012,KAWAMURA20151}, creating a spin glass phase. 
Unconventional spin glass behavior for some pyrochlore oxides such as \ymoo~\cite {Silverstein2014,Raju1992,Gingras1997,Ofer2010,Greedan2009}, 
\ymno~\cite{Reimers-Greedan1991}, Tb$_{2}$Mo$_{2}$O$_{7}$~\cite{Gaulin1992}, and A$_{2}$Sb$_{2}$O$_{7}$(A=Mn, Co, Ni)~\cite{ZHOU2010} has been observed 
despite the fact that their crystal structures are highly pure and chemically ordered. 
One of the most exciting pyrochlore spin glass is \ymoo~in which, instead of chemical disorder, 
the orbital degrees of freedom play a role in converting this material to a spin-glass system.
While for \ymoo, theoretical and experimental evidence of its spin-glass behavior is available, for \ymno, 
a very similar compound, the experimental results are inconsistent, and there is no theoretical study regarding its low-temperature magnetic ground state.

Reimers {\sl{et al.}}
showed~\cite{Reimers-Greedan1991} that \ymno~exhibits
experimental evidences of spin glass state below critical temperature
($T_\text{c}=20$~K), such as splitting of zero-field cooling and
field-cooling dc susceptibility, frequency dependent ac susceptibility
cusp, broad maximum peak in magnetic specific heat near $T_\text{c}$ and
no magnetization saturation in the presence of magnetic field even at
low temperatures. Also, neutron scattering measurements revealed that
there is no Bragg magnetic peak below $T_\text{c}$. In contrast, Shimakawa {\sl{et
    al.}}  observed~\cite{Shimakawa1999} that there is a lambda-peak
at $T_\text{c}=16$~K of the magnetic specific heat, a signature of the
long range ferromagnetic ground state similar to other members of this
oxide family (In$_{2}$Mn$_{2}$O$_{7}$, Lu$_{2}$Mn$_{2}$O$_{7}$,
Tl$_{2}$Mn$_{2}$O$_{7}$). 
In both studies, the authors claim that samples of \ymno~are highly pure and free of chemical disorders.

The motivation of the present study is to resolve the issue of the low-temperature magnetic ground state of \ymno~ 
by means of the density functional theory (DFT)  and classical Monte Carlo (MC) simulations. 
We derive an effective spin Hamiltonian for \ymno. The Hamiltonian consists of  
Heisenberg exchange, bi-quadratic (B), Dzyaloshinskii-Moriya (DM)
interactions, and single-ion anisotropy ($\Delta$).
Using this Hamiltonian, the critical and curie-Weiss temperature, as well as magnetic order, are investigated by Monte Carlo
simulations.
Our calculations show that  \ymno~is  a pyrochlore ferromagnet.
To ensure that the ground state is a ferromagnet, not a spin-glass, we investigate on $d$-orbital degrees of freedom as one of the possible sources of spin glass. 
Since \ymoo~is a spin glass (due to $d$-orbital degrees of freedom), we also study this material as a prototype of unconventional spin glass for comparison.
Our {\it ab initio} results indicate $d$ orbital degeneracy (orbital degrees of freedom) for Mo in \ymoo,
while Mn $d$ state is free of orbital degeneracy in \ymno.

In the following sections, we explain why \ymno~is a pyrochlore ferromagnet, not a spin glass.
In section~\ref{sec:methodology},
we present the details of the DFT and Monte Carlo computations. 
Section ~\ref{sec:results} is devoted to the spin Hamiltonian and critical properties of \ymno.
Furthermore, we compare \ymno~with \ymoo~in terms of orbital degeneracy for the minority spin state.
Finally, In section~\ref{sec:conc} a summary is given.
\begin{table*}
  \centering
  \caption{Calculated spin Hamiltonian constants for different $U_\text{eff}$ parameters. Negative and
    positive value for Heisenberg exchanges parameters denotes anti-ferromagnetic and ferromagnetic
    exchange interaction, respectively. The values of critical and Curie-Weiss temperature for different $U_\text{eff}$ parameters 
    have been calculated through Monte Carlo simulation.}
  \begin{tabular}{ccccccccc}
    \hline
    $U_\text{eff}$ (eV) & $J_\text{1}$ (meV) & $J_\text{2}$ (meV) & $J_\text{3a}$ (meV) &$B$(meV)&$D$(meV)&$\Delta$(meV)&$T_\text{C}$(K)&$\Theta_\text{CW}$(K)\\
    \hline
    
    2.36& -0.69 & -0.14 & -0.28& -0.79 & 0.13 & -0.12&7.0&-30.4\\
    3.36& -0.02 & -0.10 & -0.17& -0.84 & 0.13 & -0.12&7.4&-8.6\\
    4.36&  0.60 & -0.07 & -0.08& -0.89 & 0.13 & -0.12&8.3&8.6\\
    4.83&  0.89 & -0.06 & -0.05& -0.92 & 0.13 & -0.12&13.7&16.5\\
    5.36&  1.22 & -0.04 & -0.01& -0.96 & 0.14 & -0.12&21.1&25.5\\
    \hline
    Exp. & & & & & & & 20~\cite{Reimers-Greedan1991}, 16~\cite{Shimakawa1999}& 41~\cite{Reimers-Greedan1991}, 50~\cite{Shimakawa1999}\\
     \hline
    \label{tab1}
  \end{tabular}
\end{table*}
\section{Computational details}
\label{sec:methodology} 
For all of the {\it ab initio} calculations, we adopt experimental parameters for both \ymno~and \ymoo. 
These pyrochlores crystallize in a cubic structure with $Fd\overline{3}m$ space-group. 
Both Mn(Mo) and Y ions form separate corner-sharing tetrahedral networks (Fig.~\ref{fig:pyrochloe}
shows tetrahedral networks of Mn atoms in pyrochlore
structure), where Y, Mn(Mo), O(1) and O(2) atoms occupy
$16d$, $16c$, $48f$, and $8a$ Wyckoff position, respectively.
From X-ray powder diffraction, the x-positional internal parameter and lattice
constant of \ymno~is determined 0.3274(8) and 9.901  $\mathring{\mathrm{A}}$, respectively~\cite{SUBRAMANIAN1988}.
For \ymoo, powder neutron diffraction identifies internal parameter as 0.3382(1) and lattice constant as 10.230(1) $\mathring{\mathrm{A}}$~\cite{REIMERS1988}.

In this work, we use  FLURE~\cite{fleur} and Quantum-Espresso~\cite{Giannozzi_2009} (QE) code for {\it ab initio} calculations. 
The former operates based on the full-potential linearized augmented plane wave (FPLAPW), 
while the latter uses the plane-wave pseudo-potential approach. 
For FPLAPW calculations, the optimized cut-off of wave function expansion in the
interstitial region is set to
$k_{\mathrm{max}}=4.2\,\ \mathrm{a.u.}^{-1}$. The muffin-tin radius of
Y, Mn and O atoms are set to 2.6, 2.1 and 1.4 a.u., respectively. 
For all QE calculations, the optimized 40 Ry and 400 Ry cutoff have been considered for expanding wavefunction and charge density in plane wave, respectively.
In QE calculations, we approximate electron-ion interactions using the GBRV ultra-soft pseudo-potential.~\cite{Vanderbilt2014}. 
For the exchange-correlation energy, we employ the Perdew Burke Ernzerhof
parametrization of the generalized gradient approximation (GGA)~\cite{Perdew1996}.
For the Bader charge analysis, we use Critic2~\cite{critic2,critic} with QE.

We use the GGA+$U$ approximation to correct
the on-site Coulomb interaction for $3d$ and $4d$ electrons of Mn and Mo atoms, respectively.
The implementation of GGA+$U$ in FLEUR follows Liechtenstein's approach. 
The approach needs two parameters, $U$ (on-site Coulomb repulsion) and  $J_{\rm{H}}$ (the on-site Hund exchange). 
While in our QE calculations, we use Dudarev's approach, which needs only effective on-site Coulomb repulsion 
($U_\text{eff}=U-J_{\rm{H}}$). Because for many oxides $J_{\rm{H}} \sim 1$~eV~\cite{Anisimov1991,Vaugier}, we use this value for $J_{\rm{H}}$ in all of FPLAPW calculations.
To have an estimation of the $U_\text{eff}$ parameter, we employ the linear response calculation method implemented in QE~\cite{Cococcioni2005}. 
For this estimation, we use the conventional cell of \ymno~(containing 88 atoms).
We also estimate $U_\text{eff}$ through a more exact procedure called a self-consistent Hubbard $U$~\cite{Kulik2006}. 
The method calculates $U_\text{eff}$ through repeating linear response calculation within GGA+$U$ until $U_\text{eff}$ reaches to a convergent value.

We derive the spin model Hamiltonian for \ymno~ from FPLAPW total energies of different magnetic configurations.
In order to capture the physics of exchange interactions and critical properties of \ymno~pyrochlore, 
we define the spin Hamiltonian as follows:
\begin{equation}
  \label{H}
  \begin{split}
    H = -\frac{1}{2}\sum_{i\neq j}
    J_{ij}(\vec{S_{i}}\cdot\vec{S_{j}})+\frac{1}{2}B\sum_{\rm n.n}
    (\vec{S_{i}}\cdot\vec{S_{j}})^{2} \\
    +\frac{1}{2}D \sum_{\rm n.n}
    \hat{D}_{ij}\cdot(\vec{S_{i}}\times \vec{S_{j}})+\Delta\sum_{i}
    (\vec{S_{i}}\cdot\vec{d_{i}})^{2}
  \end{split}
\end{equation}
where $\vec{S_{i}}$ denotes unit vector of magnetic moment at $i$th lattice site, 
$J_{ij}$ are the Heisenberg exchange parameters up to third neighbor $(J_{1}, J_{2}, J_{3a})$
(please see Ref. \onlinecite{sadeghi2015}  for more details),
$B$ is the bi-quadratic exchange interaction between
the nearest neighbor, $D$ shows the strengths of DM
interaction and $\Delta$ is the strength of single-ion anisotropy, respectively.
Also, $\hat{D}_{ij}$ shows direction of DM vectors which is determined by Moriya rules ~\cite{Moriya1960} and 
vector $\vec{d_{i}}$ is the single-ion easy-axis direction at $i{\rm th}$ site.
The calculation of different  spin Hamiltonian terms is divided into two categories.
First category is related to the Heisenberg term up to the third nearest neighbor. To this end, we use the conventional
unit-cell (88 atoms) with $4\times4\times4$ Monkhorst-Pack k-point mesh. Second category is devoted to other spin Hamiltonian terms. 
Since the nearest neighbor is important for these terms, we consider the primitive cell (22 atoms) with $6\times6\times6$
Monkhorst-Pack k-point mesh.
The methods of calculation of the different spin Hamiltonian terms have
been reported in ~Ref.[\onlinecite{sadeghi2015}]. 

To explore the low temperature magnetic ground state and critical
properties of Y$_{2}$Mn$_{2}$O$_{7}$, we perform Monte Carlo
simulations using the replica exchange method~\cite{Hukushima1996} as
implemented in the Esfahan Spin Simulation package
(ESpinS).~\cite{rezaei2019espins} We use three-dimensional lattices
consisting of $N\times L^{3}$ spins, where $L=11$ is
the linear size of the simulation cell and $N$ is the number of spins
($N=4$ for one tetrahedron).  $10^{6}$ Monte Carlo steps (MCs) per
spin at each temperature are considered for the thermal equilibrium
and data collection, respectively. To reduce the correlation between
the successive data, measurements are done after skipping
every 10
MCs.

\section{Results and Discussion}
\label{sec:results}

\subsection{Spin Hamiltonian of \ymno}

Table.~\ref{tab1} summarizes calculated $H$ terms for different $U_\text{eff}$
parameters as well as $T_\text{C}$ and $\Theta_\text{CW}$ which have obtained by means of MC simulations of Eq.~\eqref{H}.
From the linear response and self-consistence linear response approach, we obtain $U_\text{eff}$=4.36 eV and 4.83 eV, respectively.
According to the Table.~\ref{tab1},
for these $U_\text{eff}$ values, the type of nearest neighbor exchange interaction is ferromagnetic. 
The linear response method  provides a rough estimation of 
$U_\text{eff}$, which allows us to tune $U_\text{eff}$ around this estimation to obtain more consistent results with the experiment.
Therefore we also consider $U_\text{eff}$=3.36 eV, and $U_\text{eff}$=5.36 eV.
Because at $U_\text{eff}$=3.36 eV, $J_1$ is approximately zero, we also add $U_\text{eff}$=2.36 eV in our Hubbard parameter list, where we  guess $J_1$ becomes antiferromagnetic.

According to the Table.~\ref{tab1}, the most influential interactions are the first nearest neighbor Heisenberg and bi-quadratic exchange. 
For the Heisenberg exchange, $J_{1}$, there is a transition at $U_\text{eff}$=3.36 from anti-ferromagnetic (negative $J_{1}$) to ferromagnetic (positive $J_{1}$) type.
The anisotropic exchange interactions ($D$ and $\Delta$), unlike isotropic terms ($J_{1}$ and $B$), have no changes with $U_\text{eff}$. 
The positive $D$ (direct Dzyaloshinskii-Moriya interaction) 
tends to choose noncollinear spin orientations ~\cite{Elhajal2005,Donnerer2016}
(all-in/all-out phase) but this interaction is weaker as compare as
other interactions, and the ground state is likely to be less
dependent on this interaction. 
In contrast, the bi-quadratic interaction ($B$) becomes slightly stronger with increasing $U_\text{eff}$, and its negative value indicates 
that magnetic moments tend to choose collinear spin orientation.

\subsection{Monte Carlo simulations}\label{sec:MCS}
\begin{figure}
        \includegraphics[width=0.85\columnwidth]{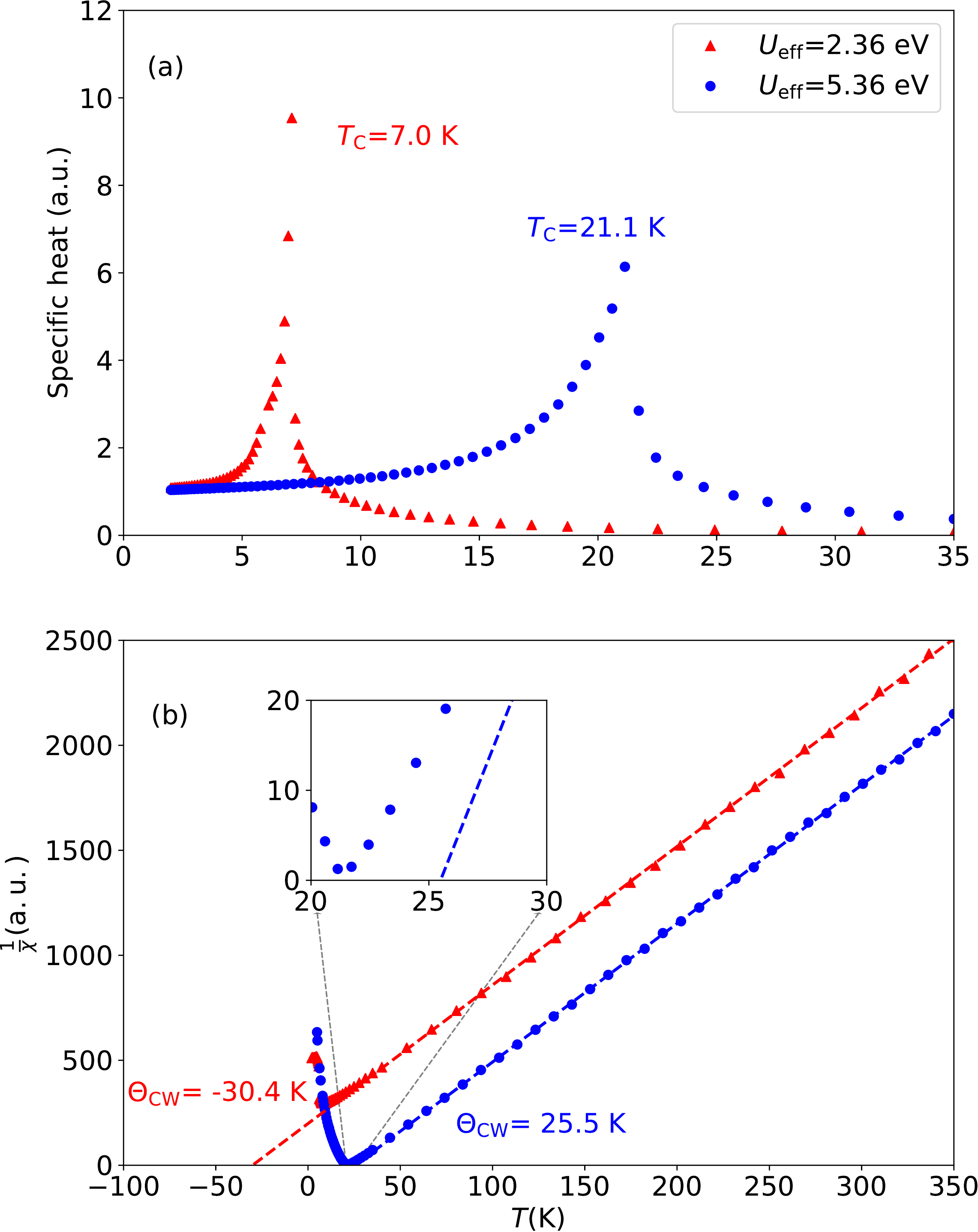}
    \caption{(Color online) Monte Carlo simulations results using a lattice of
    linear size $L=11$: a) magnetic specific heat, b) reversed magnetic susceptibility for $U_\text{eff}=2.36$ and 5.36~eV . Dashed lines are fitting lines to data in range 250-300 K.
                      The inset zoom shows reversed magnetic susceptibility in 20-30 K interval for U=5.36 eV. }
    \label{fig:MC}
\end{figure}

For each $U_{\rm{eff}}$,  the corresponding spin model Hamiltonian is used in MC simulations. From MC simulations,  we obtain
thermodynamic magnetic properties of \ymno, such as critical ($T_\text{C}$) and Curie-Wise temperature ($\Theta_\text{CW}$).
To extract $\Theta_\text{CW}$, we linearly extrapolate the reversed magnetic susceptibility data at high temperature (250-300 K) 
toward the low-temperature, as shown in Fig.~\ref{fig:MC}(b).
For estimation of $T_\text{C}$, we use the peak of magnetic specific heat (Fig.~\ref{fig:MC}(a)). Comparing $T_\text{C}$ and $\Theta_\text{CW}$ with experiment (Table~\ref{tab1}) 
reveals that the MC results related to self-consistent $U_{\rm {eff}}$ (i.e., 4.83 eV)  and also $U_{\rm {eff}}=5.36$ eV are comparable to experimental data. 
Also, according to Table~\ref{tab1}, there is a transition from antiferromagnetism ($\Theta_\text{CW} < 0$) to ferromagnetism ($\Theta_\text{CW} > 0$) 
when $U_{\rm {eff}}$ changes from 3.36 to 4.36.
So, although we find $U_{\rm {eff}}=5.36$ eV as the best of our GGA+$U$ result, 
considering two extremums of our results {\it i.e.}, $U_{\rm {eff}}=2.36$ eV (antiferromagnetic phase) and $U_{\rm {eff}}=5.36$ eV (ferromagnetic phase), can be instructive.

For $U_{\rm {eff}}=5.36$ eV, ferromagnetic $J_1$  dominates over other interactions and it is expected to have a ferromagnetic phase. 
The strong negative bi-quadratic interaction ($B=-0.96$ meV) prevents non-collinear magnetism,
which can be caused by single-ion ($\Delta=-0.12$ meV) and Dzyaloshinskii-Moriya ($D=0.14$ meV) interactions.
However, for $U_{\rm {eff}}=2.36$ eV, although $J_1$ and $B$ play 
a significant role, the third nearest neighbor exchange interaction ($J_{3a}$)
also contributes to determine the magnetic ordering configuration at low temperatures.
In the following, we consider the effect of $J_{3a}$ in the magnetic ordering of the system.

\begin{figure}
    \centering
      \includegraphics[width=0.95\columnwidth]{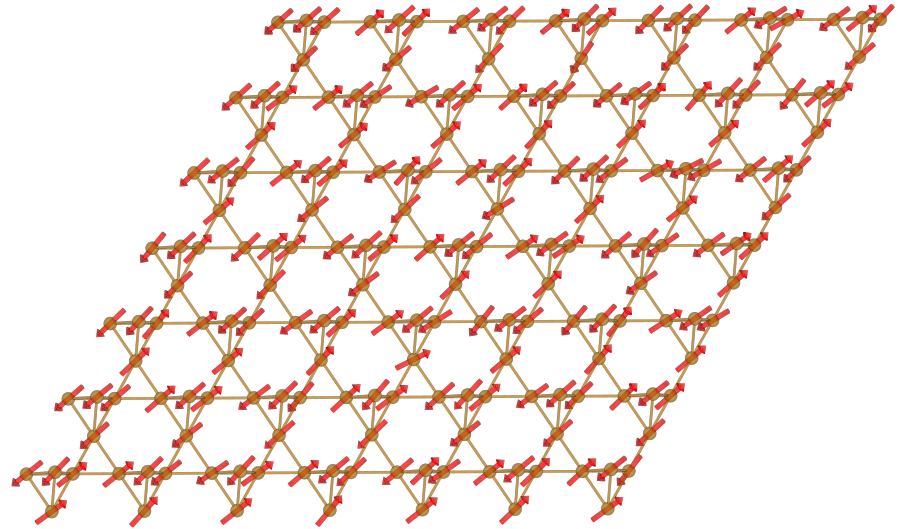}
      \caption {(Color online) A two dimensional slice ($7\time7\times7\times1$) of spin snapshot at $T=0.5$ K for $U_{\rm {eff}}=2.36$ eV }
    \label{fig:snapshot}
\end{figure}

To have a correct sense about magnetic ordering below the transition temperature for $U_{\rm {eff}}=2.36$ case, 
we take a snapshot of the magnetic configuration at 0.5 K (please see Fig.~\ref{fig:snapshot} ).
From the snapshot, we can easily recognize a collinear antiferromagnetic arrangement 
in most of the tetrahedrons. While in few of the tetrahedrons, the magnetic arrangement doesn't follow a perfect antiferromagnetic configuration 
(i.e., $|{\bf S}_1+{\bf S}_2+{\bf S}_3+{\bf S}_4|\approx0$ where ${\bf S}_1 \dots {\bf S}_4$ indicate
the unit vectors of spin directions at each vertex).   
Still, the characteristic feature in all of the tetrahedrons is that magnetic moments are approximately 
collinear.  Therefore, to consider the magnetic ordering, we guess absolute total magnetization, $|{\bf S}_1+{\bf S}_2+{\bf S}_3+{\bf S}_4|$, as 
a proper measurement for magnetic order inside tetrahedrons (for a collinear arrangement,  $|{\bf S}_1+{\bf S}_2+{\bf S}_3+{\bf S}_4|$ can be 4, 2 or 0).
\begin{figure}
    \centering
      \includegraphics[width=0.95\columnwidth]{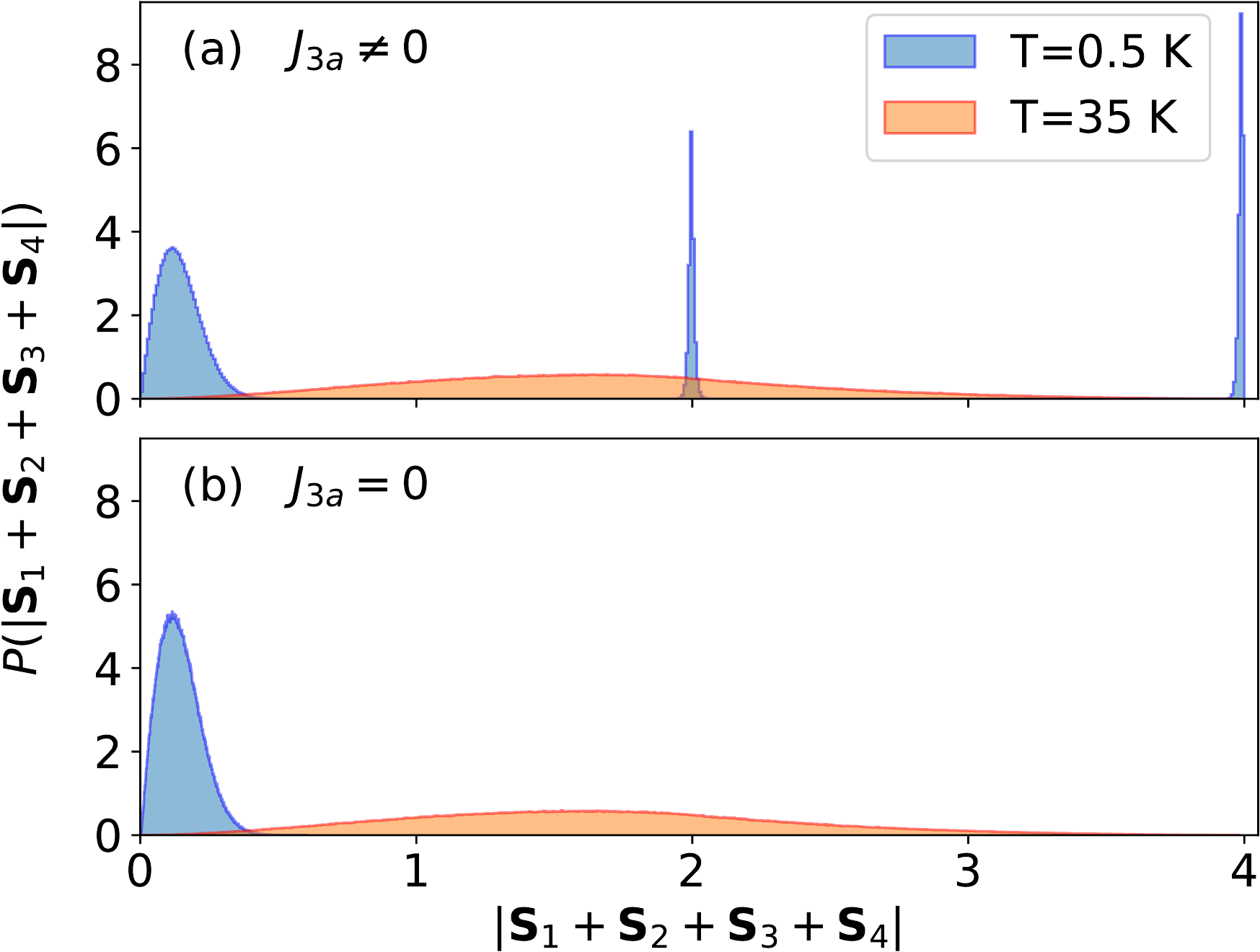}
      \caption {(Color online) The probability distribution of $|{\bf S}_1+{\bf S}_2+{\bf S}_3+{\bf S}_4|$ inside tetrahedrons at $T=0.5$ K for $U_{\rm {eff}}=2.36$ eV (a) with $J_{3a}$
                                (b) without $J_{3a}$. }
    \label{fig:pdf}
\end{figure}
The probability density of $|{\bf S}_1+{\bf S}_2+{\bf S}_3+{\bf S}_4|$ at T=0.5 K (below the transition temperature) 
and T=35 K (above the transition temperature) is represented in  Fig.~\ref{fig:pdf}(a). 
The distribution of $|{\bf S}_1+{\bf S}_2+{\bf S}_3+{\bf S}_4|$ indicates  the collinear magnetism (the peaks are around at 0, 2, and 4), 
but in the many tetrahedrons, a perfect antiferromagnetic arrangement is disturbed.
However, an almost perfect antiferromagnetic arrangement will appear (Fig.~\ref{fig:pdf}(b)) 
if we switch off the third nearest neighbor exchange interaction ($J_{3a}$).
Therefore, although our MC simulation shows $J_{3a}$ has a very small effect on transition temperature, 
but it changes completely the magnetic order.

\subsection{Orbital degree of freedom in \ymno~and \ymoo}
Our \textit {ab-initio} calculations show that for both calculated $U_\text{eff}$ using linear-response (4.36~eV) and self-consistent linear-response (4.83~eV), 
the low temperature magnetic ground state has a ferromagnetic order.
Also, MC simulations confirm the compatibility of the derived spin Hamiltonian from GGA+$U$ at $U_\text{eff}$=4.83 and 5.36~eV
with the experimental thermodynamic quantities, $T_C$ and $\Theta_\text{CW}$.
However, some experimental evidence indicate spin-glass behavior for this compound~\cite{Reimers-Greedan1991}.
Since experimental results for \ymno~confirm no chemical disorder present in this system, 
we suggest that \ymno~may have orbital degrees of freedom mechanism similar to \ymoo~which leads them to a spin glass behavior~\cite{Silverstein2014,Shinaoka2013}.
In the following, we compare \ymno~and \ymoo~ to confirm firmly that \ymno~ can not be a spin-glass similar to \ymoo. 

\subsubsection{Electronic structure differences between \ymno~and \ymoo~at GGA level} 
\begin{figure}
    \centering
    \includegraphics[width=0.90\columnwidth]{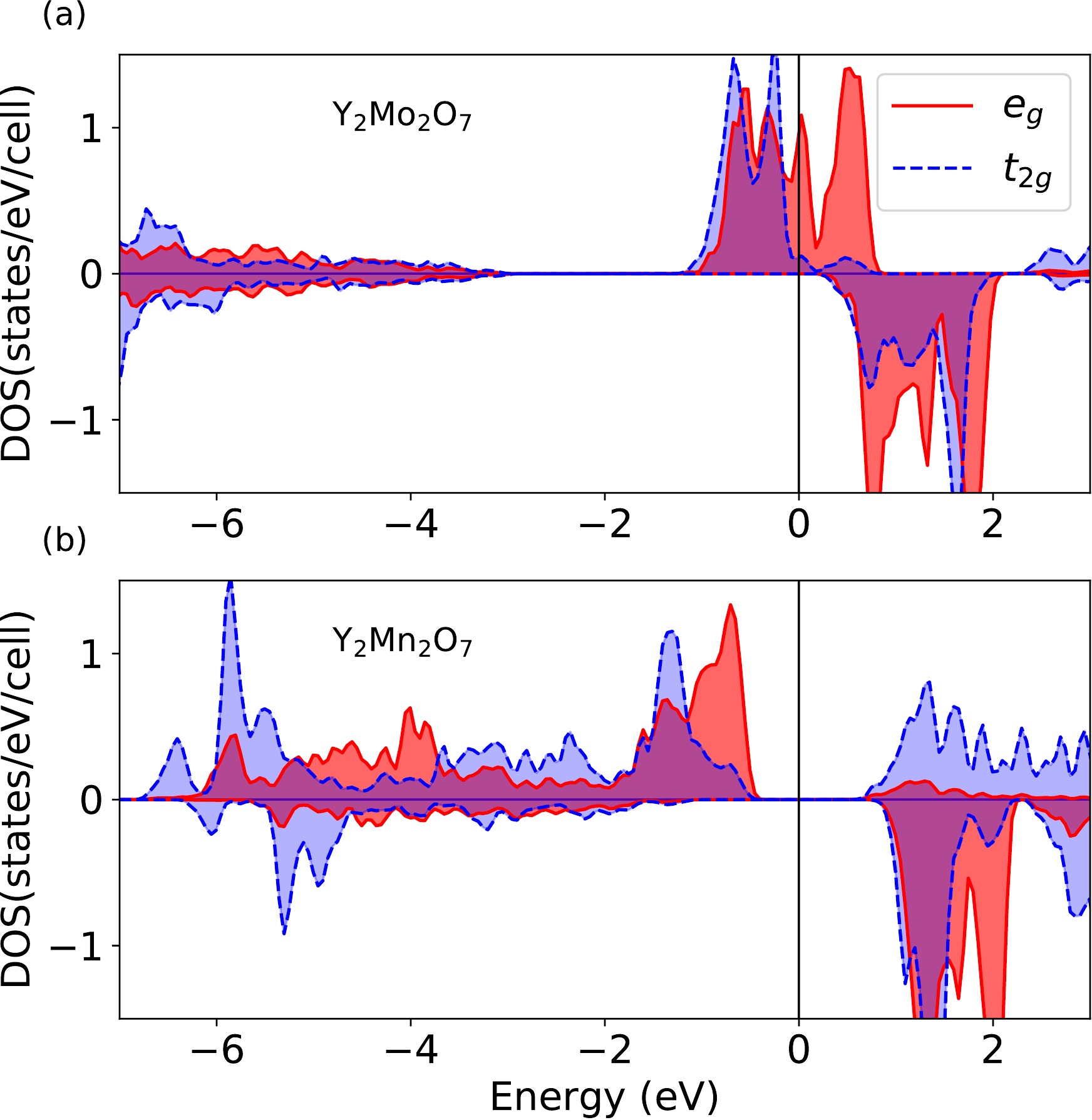} 
    \caption{(Color online) density of state (DOS) of (a) Mo $d$-orbital in \ymoo (b), and Mn $d$-orbital in \ymno. 
             The positive and negative DOS indicates majority-spin and minority-spin states, respectively.   }
    \label{fig:pdos}
\end{figure}

In the literature of magnetic pyrochlore oxides~\cite{Gardner2010}, 
the valence state for pyrochlore oxides is simplified as A$_{2}^{+3}$B$_{2}^{+4}$O$_7^{-2}$. 
This simplification leads to $3d^3$ and $4d^2$ electron configurations in $d$ orbitals for \ymno~and \ymoo, respectively~\cite{Reimers-Greedan1991,Shinaoka2013}.
However, using the Bader charge analysis, we estimate the valence state within the GGA calculations as follows: Y$_{2}^{+2.18}$Mn$_{2}^{+1.89}$O(1)$_5^{-1.12}$O(2)$_2^{-1.39}$
Y$_{2}^{+2.20}$Mo$_{2}^{+2.19}$O(1)$_5^{-1.23}$O(2)$_2^{-1.40}$. 
Lowdin charge analysis also confirms these results. 
In addition, Lowdin charge analysis indicates how electrons distribute among $d$ orbitals (Table ~\ref{tab:d-orb}). 
Due to the hybridization of the $d$-orbitals of Mn (Mo) with $p$-orbitals of oxygen atoms, 
the occupations of $d$-orbitals of Mn (Mo) is fractional. The spin-minority occupation
for both Mn and Mo is about 1 electron, while the spin-majority occupation is about 4 and 3 for Mn and Mo receptively.
To have better comparison, we also compare density of state (DOS) of $d$-Mo and $d$-Mn states at the GGA level of DFT theory (Fig.~\ref{fig:pdos}).
According to Fig.~\ref{fig:pdos}, \ymno~is an insulator indicating that band theory gives at least a correct state (i.e. insulating) for this material.
However, DOS of \ymoo~shows some $d$-states at Fermi level despite the fact that \ymoo~ is a semiconductor~\cite{Subramanian1980}.
The reason for such a difference is that while crystal field splitting helps $3d^5$ configuration of \ymno~becomes an insulator,
the lack of crystal field splitting at the Fermi level leads $4d^4$ configuration of \ymoo~to the metal in GGA calculations.


\begin{table}
\caption{ Charge distribution among Mn-$d$ and Mo-$d$ orbitals in \ymno~ and \ymoo, respectively, obtained by GGA Lowdin charge analysis.}
\label{tab:d-orb}
\begin{ruledtabular}
\begin{tabular}{c|cccc}
atom             & spin          &  $d_{tot}$   & $(d_{z^2}, d_{x^2-y^2})$   & $(d_{xy}, d_{xz}, d_{yz})$  \\ \hline
\multirow{2}{*}{Mn} & $\uparrow$        & $3.9589$     & $0.9450$  &      $0.6897$           \\
                      & $\downarrow$    & $1.1277$     & $0.1819$  &      $0.2546$    \\ \hline
\multirow{2}{*}{Mo} & $\uparrow$        & $3.0019$     & $0.6326$  &      $0.5789$           \\
                      & $\downarrow$    & $1.1764$     & $0.1923$  &      $0.2639$    \\
\end{tabular}
\end{ruledtabular}
\end{table}

\subsubsection{Orbital degree of freedom}
The GGA+$U$ calculations are generally based on the density matrix of atomic orbital states such as $d$-states. 
In some systems, there are several choices for density-matrix occupations and therefore there are several electronic structures for the system~\cite{Meredig2010,DEILYNAZAR2015}
(more precisely, GGA+$U$ faces multi-minima problem).
In practice, the correct density-matrix occupation can be chosen by comparing the GGA+$U$ total energies.
Principally, for systems such as \ymno~in which GGA predicts correctly the system as an insulator, 
there is a unique density-matrix occupation ({\it i.e.}, a single minimum) 
, and there is no need to optimize density-matrix.
However, for systems such as \ymoo~in which GGA results in a (wrong) metallic state, 
it is possible to have multi-minima due to degrees of freedom in density-matrix occupation. 
These degrees of freedom in the density matrix happens because of partially occupied $d$-orbital states at the Fermi level.
It is also worth mentioning that the density-matrix occupations are under the influence of symmetry. 
Higher symmetry lowers degrees of freedom in density matrix, 
and in some cases such as \ymoo, restricts GGA+$U$ calculations to some wrong solutions.

\begin{figure} 
  \centering
  \includegraphics[width=0.95\columnwidth]{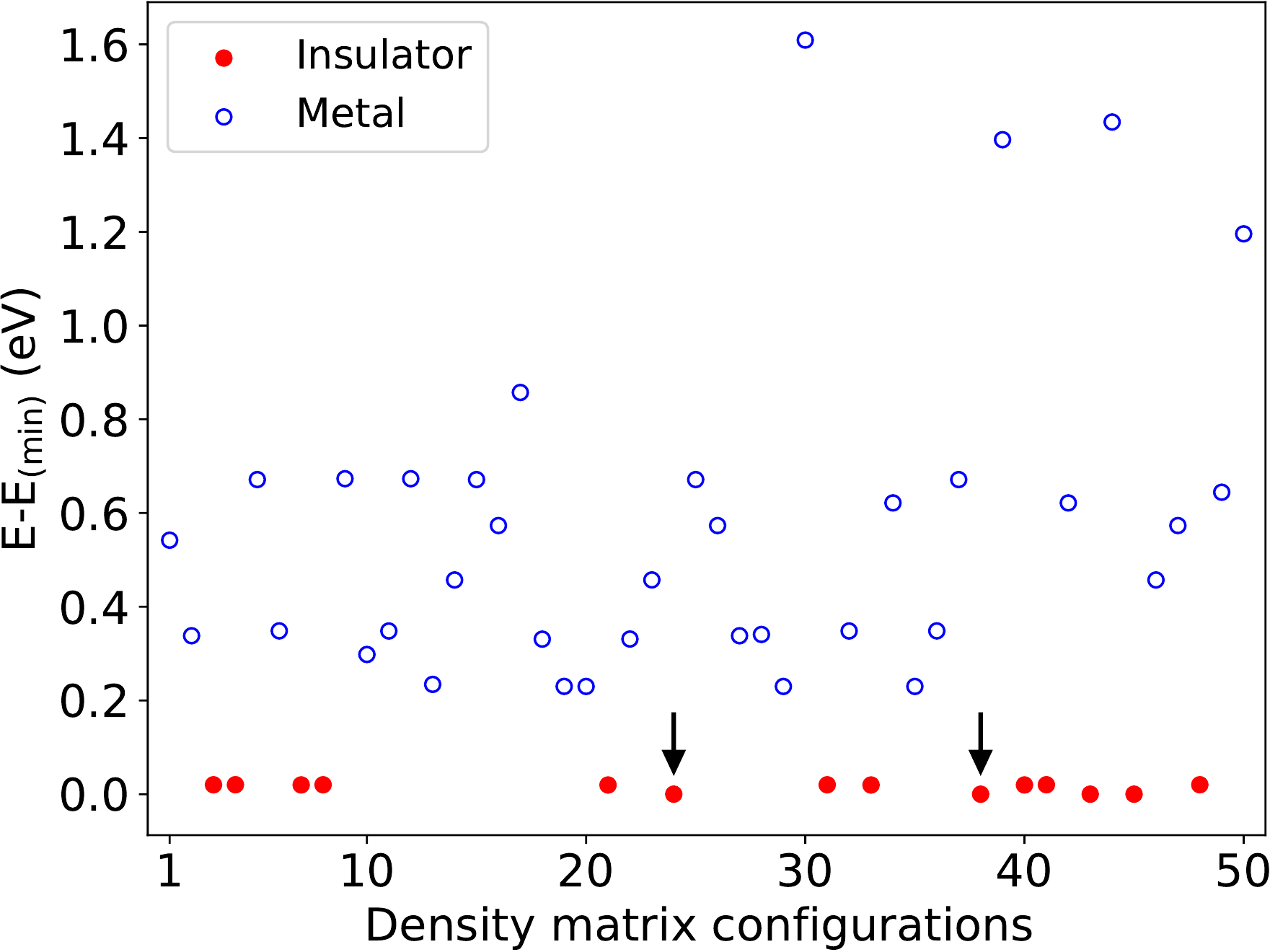}
  \caption{Total energy of \ymoo~versus density matrix occupations. 
           The x-axis representatively indicates different density matrix occupations. 
           We set the minimum of total energy  to zero. The red and unfilled blue circles indicate insulator and metallic solutions, respectively. 
           The black arrows show two insulator solutions that we select them as instances to examine $d$ orbital orientation.}
  \label{fig:orbital_degeneracy}
\end{figure}

\begin{figure} 
  \centering
  \includegraphics[width=0.95\columnwidth]{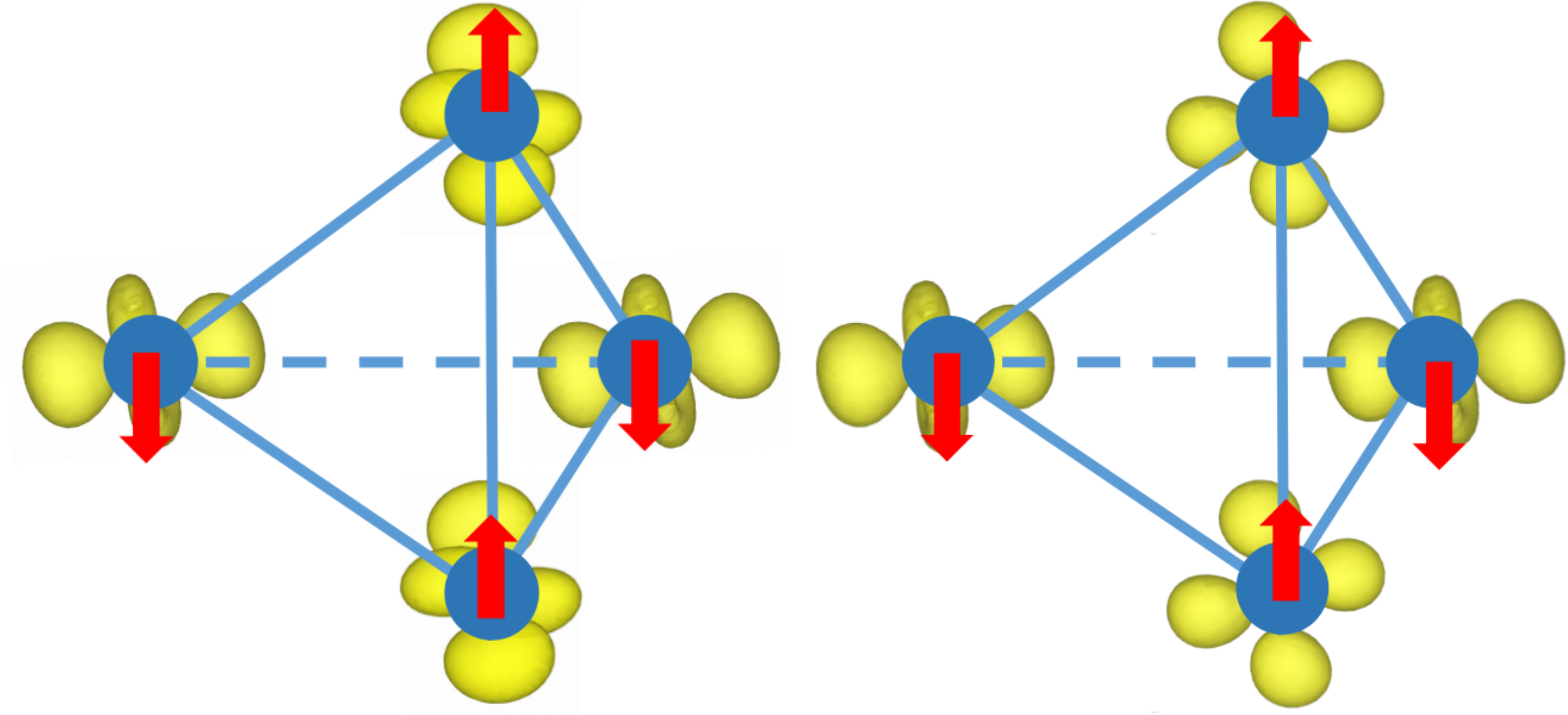}
  \caption{The orbital orientation of 4$d$-band nearly below Fermi level for two approximately degenerate GGA+$U$ solutions for \ymoo}
  \label{fig:oo}
\end{figure}

To explore the possibilities in density-matrix occupation for \ymoo, 
we manipulate the initial of density-matrix occupations of Mo atoms at the starting point of GGA+$U$ calculations with symmetry breaking 
for the antiferromagnetic magnetic configuration.
For our exploration in density-matrix occupation, we consider 150 different initial density-matrix occupations.
These initial density-matrix occupations lead to 50 distinguishable electronic configurations as the results of GGA+$U$  self-consistent field (SCF) calculations.
Fig.~\ref{fig:orbital_degeneracy} shows the total energy of these 50 GGA+$U$ solutions versus their density matrix occupations, 
which is represented by a number (from 1-50).
Among these 50 GGA+$U$ solutions, we focus on insulating ones. 
Despite differences in density-matrix occupations, the insulating GGA+$U$ solutions are nearly degenerate in terms of energy. 
To characterize the difference between these nearly degenerate states, we select two of them (the ones pointed by arrows in Fig.~\ref{fig:orbital_degeneracy}) 
and compare their 
$d$ orbital orientations at the tetrahedron corners.
Fig.~\ref{fig:oo} shows that despite the same spin magnetic moment directions, there are tiny differences in orbital orientations 
of these two electronic configurations. 
This orbital degrees of freedom create local distortions and  causing the system to show spin-glass behavior~\cite{Silverstein2014,Shinaoka2013}.
In contrast, for \ymno~, we reach to a single solution where orbital orientations at the tetrahedron corners show a single direction.
\section{Conclusions}\label{sec:conc}
In this paper, we tried to shed light on the magnetic state of pyrochlore \ymno~ using ab initio GGA+$U$ calculations and Monte Carlo simulations.
For GGA+U calculations, we estimated Hubbard $U_\text{eff}$ via the linear response method.
Using GGA+$U$ calculations, we constructed several spin models for \ymno~ by tuning Hubbard parameter $U_\text{eff}$ around its estimated value.
By comparing Monte Carlo simulations of these models with experimental measurements such as Curie-Weiss temperature,  
we found that the ferromagnetic state of \ymno~ matches almost to the experiments.
Also, we proved that \ymno~ cannot have $d$ orbital degrees of freedom mechanism which may 
turn \ymno~ into a spin glass system. To examine this issue, we analyzed both \ymno~ and \ymoo~, 
where for the latter, there are evidences of spin glass behavior due to $d$ orbital degrees of freedom mechanism.
We demonstrated that while GGA+$U$ calculations reveal $d$ orbital degrees of freedom mechanism in \ymoo, 
there is no way that  GGA+$U$ leads to such machinery for \ymno. 
In summary, we suggest that a pure pyrochlore phase of \ymno~ is a simple ferromagnetic system. 
We also propose further experiments on this material to clarify this issue.

\begin{acknowledgments}
M. Amirabbasi thanks Marjana Lez\v{a}i\'c, Gustav Bihlmayer and Farhad Shahbazi for useful discussions.
M. Alaei thanks Nafise Rezaei for her constructive comments.
\end{acknowledgments}

\bibliographystyle{apsrev4-1}
\bibliography{bib}

\begin{thebibliography}{40}%
\makeatletter
\providecommand \@ifxundefined [1]{%
 \@ifx{#1\undefined}
}%
\providecommand \@ifnum [1]{%
 \ifnum #1\expandafter \@firstoftwo
 \else \expandafter \@secondoftwo
 \fi
}%
\providecommand \@ifx [1]{%
 \ifx #1\expandafter \@firstoftwo
 \else \expandafter \@secondoftwo
 \fi
}%
\providecommand \natexlab [1]{#1}%
\providecommand \enquote  [1]{``#1''}%
\providecommand \bibnamefont  [1]{#1}%
\providecommand \bibfnamefont [1]{#1}%
\providecommand \citenamefont [1]{#1}%
\providecommand \href@noop [0]{\@secondoftwo}%
\providecommand \href [0]{\begingroup \@sanitize@url \@href}%
\providecommand \@href[1]{\@@startlink{#1}\@@href}%
\providecommand \@@href[1]{\endgroup#1\@@endlink}%
\providecommand \@sanitize@url [0]{\catcode `\\12\catcode `\$12\catcode
  `\&12\catcode `\#12\catcode `\^12\catcode `\_12\catcode `\%12\relax}%
\providecommand \@@startlink[1]{}%
\providecommand \@@endlink[0]{}%
\providecommand \url  [0]{\begingroup\@sanitize@url \@url }%
\providecommand \@url [1]{\endgroup\@href {#1}{\urlprefix }}%
\providecommand \urlprefix  [0]{URL }%
\providecommand \Eprint [0]{\href }%
\providecommand \doibase [0]{http://dx.doi.org/}%
\providecommand \selectlanguage [0]{\@gobble}%
\providecommand \bibinfo  [0]{\@secondoftwo}%
\providecommand \bibfield  [0]{\@secondoftwo}%
\providecommand \translation [1]{[#1]}%
\providecommand \BibitemOpen [0]{}%
\providecommand \bibitemStop [0]{}%
\providecommand \bibitemNoStop [0]{.\EOS\space}%
\providecommand \EOS [0]{\spacefactor3000\relax}%
\providecommand \BibitemShut  [1]{\csname bibitem#1\endcsname}%
\let\auto@bib@innerbib\@empty
\bibitem [{\citenamefont {Lacroix}\ \emph {et~al.}(2011)\citenamefont
  {Lacroix}, \citenamefont {Mendels},\ and\ \citenamefont {Mila}}]{Lacroix}%
  \BibitemOpen
  \bibfield  {author} {\bibinfo {author} {\bibfnamefont {C.}~\bibnamefont
  {Lacroix}}, \bibinfo {author} {\bibfnamefont {P.}~\bibnamefont {Mendels}}, \
  and\ \bibinfo {author} {\bibfnamefont {F.}~\bibnamefont {Mila}},\ }\href@noop
  {} {\emph {\bibinfo {title} {Introduction to Frustrated Magnetism}}}\
  (\bibinfo  {publisher} {Springer Series in Solid-State Science (Springer,
  Heidelberg)},\ \bibinfo {year} {2011})\BibitemShut {NoStop}%
\bibitem [{\citenamefont {Greedan}(2001)}]{Greedan2001}%
  \BibitemOpen
  \bibfield  {author} {\bibinfo {author} {\bibfnamefont {J.~E.}\ \bibnamefont
  {Greedan}},\ }\href@noop {} {\bibfield  {journal} {\bibinfo  {journal} {J.
  Mater. Chem}\ }\textbf {\bibinfo {volume} {11}} (\bibinfo {year}
  {2001})}\BibitemShut {NoStop}%
\bibitem [{\citenamefont {Gardner}\ \emph {et~al.}(2010)\citenamefont
  {Gardner}, \citenamefont {Gingras},\ and\ \citenamefont
  {Greedan}}]{Gardner2010}%
  \BibitemOpen
  \bibfield  {author} {\bibinfo {author} {\bibfnamefont {J.~S.}\ \bibnamefont
  {Gardner}}, \bibinfo {author} {\bibfnamefont {M.~J.~P.}\ \bibnamefont
  {Gingras}}, \ and\ \bibinfo {author} {\bibfnamefont {J.~E.}\ \bibnamefont
  {Greedan}},\ }\href@noop {} {\bibfield  {journal} {\bibinfo  {journal} {Rev.
  Mod. Phys.}\ }\textbf {\bibinfo {volume} {82}},\ \bibinfo {pages} {53}
  (\bibinfo {year} {2010})}\BibitemShut {NoStop}%
\bibitem [{\citenamefont {Savary}\ and\ \citenamefont
  {Balents}(2017)}]{Savary2017}%
  \BibitemOpen
  \bibfield  {author} {\bibinfo {author} {\bibfnamefont {L.}~\bibnamefont
  {Savary}}\ and\ \bibinfo {author} {\bibfnamefont {L.}~\bibnamefont
  {Balents}},\ }\href@noop {} {\bibfield  {journal} {\bibinfo  {journal} {Phys.
  Rev. Lett.}\ }\textbf {\bibinfo {volume} {118}},\ \bibinfo {pages} {087203}
  (\bibinfo {year} {2017})}\BibitemShut {NoStop}%
\bibitem [{\citenamefont {Gaudet}\ \emph {et~al.}(2019)\citenamefont {Gaudet},
  \citenamefont {Smith}, \citenamefont {Dudemaine}, \citenamefont {Beare},
  \citenamefont {Buhariwalla}, \citenamefont {Butch}, \citenamefont {Stone},
  \citenamefont {Kolesnikov}, \citenamefont {Xu}, \citenamefont {Yahne},
  \citenamefont {Ross}, \citenamefont {Marjerrison}, \citenamefont {Garrett},
  \citenamefont {Luke}, \citenamefont {Bianchi},\ and\ \citenamefont
  {Gaulin}}]{Gaudet2019}%
  \BibitemOpen
  \bibfield  {author} {\bibinfo {author} {\bibfnamefont {J.}~\bibnamefont
  {Gaudet}}, \bibinfo {author} {\bibfnamefont {E.~M.}\ \bibnamefont {Smith}},
  \bibinfo {author} {\bibfnamefont {J.}~\bibnamefont {Dudemaine}}, \bibinfo
  {author} {\bibfnamefont {J.}~\bibnamefont {Beare}}, \bibinfo {author}
  {\bibfnamefont {C.~R.~C.}\ \bibnamefont {Buhariwalla}}, \bibinfo {author}
  {\bibfnamefont {N.~P.}\ \bibnamefont {Butch}}, \bibinfo {author}
  {\bibfnamefont {M.~B.}\ \bibnamefont {Stone}}, \bibinfo {author}
  {\bibfnamefont {A.~I.}\ \bibnamefont {Kolesnikov}}, \bibinfo {author}
  {\bibfnamefont {G.}~\bibnamefont {Xu}}, \bibinfo {author} {\bibfnamefont
  {D.~R.}\ \bibnamefont {Yahne}}, \bibinfo {author} {\bibfnamefont {K.~A.}\
  \bibnamefont {Ross}}, \bibinfo {author} {\bibfnamefont {C.~A.}\ \bibnamefont
  {Marjerrison}}, \bibinfo {author} {\bibfnamefont {J.~D.}\ \bibnamefont
  {Garrett}}, \bibinfo {author} {\bibfnamefont {G.~M.}\ \bibnamefont {Luke}},
  \bibinfo {author} {\bibfnamefont {A.~D.}\ \bibnamefont {Bianchi}}, \ and\
  \bibinfo {author} {\bibfnamefont {B.~D.}\ \bibnamefont {Gaulin}},\
  }\href@noop {} {\bibfield  {journal} {\bibinfo  {journal} {Phys. Rev. Lett.}\
  }\textbf {\bibinfo {volume} {122}},\ \bibinfo {pages} {187201} (\bibinfo
  {year} {2019})}\BibitemShut {NoStop}%
\bibitem [{\citenamefont {Balents}(2010)}]{Balents2010}%
  \BibitemOpen
  \bibfield  {author} {\bibinfo {author} {\bibfnamefont {L.}~\bibnamefont
  {Balents}},\ }\href@noop {} {\bibfield  {journal} {\bibinfo  {journal}
  {Nature}\ }\textbf {\bibinfo {volume} {464}},\ \bibinfo {pages} {199}
  (\bibinfo {year} {2010})}\BibitemShut {NoStop}%
\bibitem [{\citenamefont {Davies}\ \emph {et~al.}(2019)\citenamefont {Davies},
  \citenamefont {Topping}, \citenamefont {Jacobsen}, \citenamefont {Princep},
  \citenamefont {Kirschner}, \citenamefont {Rahn}, \citenamefont {Bristow},
  \citenamefont {Vale}, \citenamefont {da~Silva}, \citenamefont {Baker},
  \citenamefont {Sahle}, \citenamefont {Guo}, \citenamefont {Yan},
  \citenamefont {Shi}, \citenamefont {Blundell}, \citenamefont {McMorrow},\
  and\ \citenamefont {Boothroyd}}]{Davies2019}%
  \BibitemOpen
  \bibfield  {author} {\bibinfo {author} {\bibfnamefont {N.~R.}\ \bibnamefont
  {Davies}}, \bibinfo {author} {\bibfnamefont {C.~V.}\ \bibnamefont {Topping}},
  \bibinfo {author} {\bibfnamefont {H.}~\bibnamefont {Jacobsen}}, \bibinfo
  {author} {\bibfnamefont {A.~J.}\ \bibnamefont {Princep}}, \bibinfo {author}
  {\bibfnamefont {F.~K.~K.}\ \bibnamefont {Kirschner}}, \bibinfo {author}
  {\bibfnamefont {M.~C.}\ \bibnamefont {Rahn}}, \bibinfo {author}
  {\bibfnamefont {M.}~\bibnamefont {Bristow}}, \bibinfo {author} {\bibfnamefont
  {J.~G.}\ \bibnamefont {Vale}}, \bibinfo {author} {\bibfnamefont
  {I.}~\bibnamefont {da~Silva}}, \bibinfo {author} {\bibfnamefont {P.~J.}\
  \bibnamefont {Baker}}, \bibinfo {author} {\bibfnamefont {C.~J.}\ \bibnamefont
  {Sahle}}, \bibinfo {author} {\bibfnamefont {Y.-F.}\ \bibnamefont {Guo}},
  \bibinfo {author} {\bibfnamefont {D.-Y.}\ \bibnamefont {Yan}}, \bibinfo
  {author} {\bibfnamefont {Y.-G.}\ \bibnamefont {Shi}}, \bibinfo {author}
  {\bibfnamefont {S.~J.}\ \bibnamefont {Blundell}}, \bibinfo {author}
  {\bibfnamefont {D.~F.}\ \bibnamefont {McMorrow}}, \ and\ \bibinfo {author}
  {\bibfnamefont {A.~T.}\ \bibnamefont {Boothroyd}},\ }\href@noop {} {\bibfield
   {journal} {\bibinfo  {journal} {Phys. Rev. B}\ }\textbf {\bibinfo {volume}
  {99}},\ \bibinfo {pages} {174442} (\bibinfo {year} {2019})}\BibitemShut
  {NoStop}%
\bibitem [{\citenamefont {Singh}\ and\ \citenamefont {Lee}(2012)}]{Singh2012}%
  \BibitemOpen
  \bibfield  {author} {\bibinfo {author} {\bibfnamefont {D.~K.}\ \bibnamefont
  {Singh}}\ and\ \bibinfo {author} {\bibfnamefont {Y.~S.}\ \bibnamefont
  {Lee}},\ }\href@noop {} {\bibfield  {journal} {\bibinfo  {journal} {Phys.
  Rev. Lett.}\ }\textbf {\bibinfo {volume} {109}},\ \bibinfo {pages} {247201}
  (\bibinfo {year} {2012})}\BibitemShut {NoStop}%
\bibitem [{\citenamefont {Kawamura}\ and\ \citenamefont
  {Taniguchi}(2015)}]{KAWAMURA20151}%
  \BibitemOpen
  \bibfield  {author} {\bibinfo {author} {\bibfnamefont {H.}~\bibnamefont
  {Kawamura}}\ and\ \bibinfo {author} {\bibfnamefont {T.}~\bibnamefont
  {Taniguchi}}\ }(\bibinfo  {publisher} {Elsevier},\ \bibinfo {year} {2015})\
  pp.\ \bibinfo {pages} {1 -- 137}\BibitemShut {NoStop}%
\bibitem [{\citenamefont {Silverstein}\ \emph {et~al.}(2014)\citenamefont
  {Silverstein}, \citenamefont {Fritsch}, \citenamefont {Flicker},
  \citenamefont {Hallas}, \citenamefont {Gardner}, \citenamefont {Qiu},
  \citenamefont {Ehlers}, \citenamefont {Savici}, \citenamefont {Yamani},
  \citenamefont {Ross}, \citenamefont {Gaulin}, \citenamefont {Gingras},
  \citenamefont {Paddison}, \citenamefont {Foyevtsova}, \citenamefont
  {Valenti}, \citenamefont {Hawthorne}, \citenamefont {Wiebe},\ and\
  \citenamefont {Zhou}}]{Silverstein2014}%
  \BibitemOpen
  \bibfield  {author} {\bibinfo {author} {\bibfnamefont {H.~J.}\ \bibnamefont
  {Silverstein}}, \bibinfo {author} {\bibfnamefont {K.}~\bibnamefont
  {Fritsch}}, \bibinfo {author} {\bibfnamefont {F.}~\bibnamefont {Flicker}},
  \bibinfo {author} {\bibfnamefont {A.~M.}\ \bibnamefont {Hallas}}, \bibinfo
  {author} {\bibfnamefont {J.~S.}\ \bibnamefont {Gardner}}, \bibinfo {author}
  {\bibfnamefont {Y.}~\bibnamefont {Qiu}}, \bibinfo {author} {\bibfnamefont
  {G.}~\bibnamefont {Ehlers}}, \bibinfo {author} {\bibfnamefont {A.~T.}\
  \bibnamefont {Savici}}, \bibinfo {author} {\bibfnamefont {Z.}~\bibnamefont
  {Yamani}}, \bibinfo {author} {\bibfnamefont {K.~A.}\ \bibnamefont {Ross}},
  \bibinfo {author} {\bibfnamefont {B.~D.}\ \bibnamefont {Gaulin}}, \bibinfo
  {author} {\bibfnamefont {M.~J.~P.}\ \bibnamefont {Gingras}}, \bibinfo
  {author} {\bibfnamefont {J.~A.~M.}\ \bibnamefont {Paddison}}, \bibinfo
  {author} {\bibfnamefont {K.}~\bibnamefont {Foyevtsova}}, \bibinfo {author}
  {\bibfnamefont {R.}~\bibnamefont {Valenti}}, \bibinfo {author} {\bibfnamefont
  {F.}~\bibnamefont {Hawthorne}}, \bibinfo {author} {\bibfnamefont {C.~R.}\
  \bibnamefont {Wiebe}}, \ and\ \bibinfo {author} {\bibfnamefont {H.~D.}\
  \bibnamefont {Zhou}},\ }\href@noop {} {\bibfield  {journal} {\bibinfo
  {journal} {Phys. Rev. B}\ }\textbf {\bibinfo {volume} {89}},\ \bibinfo
  {pages} {054433} (\bibinfo {year} {2014})}\BibitemShut {NoStop}%
\bibitem [{\citenamefont {Raju}\ \emph {et~al.}(1992)\citenamefont {Raju},
  \citenamefont {Gmelin},\ and\ \citenamefont {Kremer}}]{Raju1992}%
  \BibitemOpen
  \bibfield  {author} {\bibinfo {author} {\bibfnamefont {N.~P.}\ \bibnamefont
  {Raju}}, \bibinfo {author} {\bibfnamefont {E.}~\bibnamefont {Gmelin}}, \ and\
  \bibinfo {author} {\bibfnamefont {R.~K.}\ \bibnamefont {Kremer}},\
  }\href@noop {} {\bibfield  {journal} {\bibinfo  {journal} {Phys. Rev. B}\
  }\textbf {\bibinfo {volume} {46}},\ \bibinfo {pages} {5405} (\bibinfo {year}
  {1992})}\BibitemShut {NoStop}%
\bibitem [{\citenamefont {Gingras}\ \emph {et~al.}(1997)\citenamefont
  {Gingras}, \citenamefont {Stager}, \citenamefont {Raju}, \citenamefont
  {Gaulin},\ and\ \citenamefont {Greedan}}]{Gingras1997}%
  \BibitemOpen
  \bibfield  {author} {\bibinfo {author} {\bibfnamefont {M.~J.~P.}\
  \bibnamefont {Gingras}}, \bibinfo {author} {\bibfnamefont {C.~V.}\
  \bibnamefont {Stager}}, \bibinfo {author} {\bibfnamefont {N.~P.}\
  \bibnamefont {Raju}}, \bibinfo {author} {\bibfnamefont {B.~D.}\ \bibnamefont
  {Gaulin}}, \ and\ \bibinfo {author} {\bibfnamefont {J.~E.}\ \bibnamefont
  {Greedan}},\ }\href@noop {} {\bibfield  {journal} {\bibinfo  {journal} {Phys.
  Rev. Lett.}\ }\textbf {\bibinfo {volume} {78}},\ \bibinfo {pages} {947}
  (\bibinfo {year} {1997})}\BibitemShut {NoStop}%
\bibitem [{\citenamefont {Ofer}\ \emph {et~al.}(2010)\citenamefont {Ofer},
  \citenamefont {Keren}, \citenamefont {Gardner}, \citenamefont {Ren},\ and\
  \citenamefont {MacFarlane}}]{Ofer2010}%
  \BibitemOpen
  \bibfield  {author} {\bibinfo {author} {\bibfnamefont {O.}~\bibnamefont
  {Ofer}}, \bibinfo {author} {\bibfnamefont {A.}~\bibnamefont {Keren}},
  \bibinfo {author} {\bibfnamefont {J.~S.}\ \bibnamefont {Gardner}}, \bibinfo
  {author} {\bibfnamefont {Y.}~\bibnamefont {Ren}}, \ and\ \bibinfo {author}
  {\bibfnamefont {W.~A.}\ \bibnamefont {MacFarlane}},\ }\href@noop {}
  {\bibfield  {journal} {\bibinfo  {journal} {Phys. Rev. B}\ }\textbf {\bibinfo
  {volume} {82}},\ \bibinfo {pages} {092403} (\bibinfo {year}
  {2010})}\BibitemShut {NoStop}%
\bibitem [{\citenamefont {Greedan}\ \emph {et~al.}(2009)\citenamefont
  {Greedan}, \citenamefont {Gout}, \citenamefont {Lozano-Gorrin}, \citenamefont
  {Derahkshan}, \citenamefont {Proffen}, \citenamefont {Kim}, \citenamefont
  {Bo\ifmmode~\check{z}\else \v{z}\fi{}in},\ and\ \citenamefont
  {Billinge}}]{Greedan2009}%
  \BibitemOpen
  \bibfield  {author} {\bibinfo {author} {\bibfnamefont {J.~E.}\ \bibnamefont
  {Greedan}}, \bibinfo {author} {\bibfnamefont {D.}~\bibnamefont {Gout}},
  \bibinfo {author} {\bibfnamefont {A.~D.}\ \bibnamefont {Lozano-Gorrin}},
  \bibinfo {author} {\bibfnamefont {S.}~\bibnamefont {Derahkshan}}, \bibinfo
  {author} {\bibfnamefont {T.}~\bibnamefont {Proffen}}, \bibinfo {author}
  {\bibfnamefont {H.-J.}\ \bibnamefont {Kim}}, \bibinfo {author} {\bibfnamefont
  {E.}~\bibnamefont {Bo\ifmmode~\check{z}\else \v{z}\fi{}in}}, \ and\ \bibinfo
  {author} {\bibfnamefont {S.~J.~L.}\ \bibnamefont {Billinge}},\ }\href@noop {}
  {\bibfield  {journal} {\bibinfo  {journal} {Phys. Rev. B}\ }\textbf {\bibinfo
  {volume} {79}},\ \bibinfo {pages} {014427} (\bibinfo {year}
  {2009})}\BibitemShut {NoStop}%
\bibitem [{\citenamefont {Reimers}\ \emph {et~al.}(1991)\citenamefont
  {Reimers}, \citenamefont {Greedan}, \citenamefont {Kremer}, \citenamefont
  {Gmelin},\ and\ \citenamefont {Subramanian}}]{Reimers-Greedan1991}%
  \BibitemOpen
  \bibfield  {author} {\bibinfo {author} {\bibfnamefont {J.~N.}\ \bibnamefont
  {Reimers}}, \bibinfo {author} {\bibfnamefont {J.~E.}\ \bibnamefont
  {Greedan}}, \bibinfo {author} {\bibfnamefont {R.~K.}\ \bibnamefont {Kremer}},
  \bibinfo {author} {\bibfnamefont {E.}~\bibnamefont {Gmelin}}, \ and\ \bibinfo
  {author} {\bibfnamefont {M.~A.}\ \bibnamefont {Subramanian}},\ }\href@noop {}
  {\bibfield  {journal} {\bibinfo  {journal} {Phys. Rev. B}\ }\textbf {\bibinfo
  {volume} {43}},\ \bibinfo {pages} {3387} (\bibinfo {year}
  {1991})}\BibitemShut {NoStop}%
\bibitem [{\citenamefont {Gaulin}\ \emph {et~al.}(1992)\citenamefont {Gaulin},
  \citenamefont {Reimers}, \citenamefont {Mason}, \citenamefont {Greedan},\
  and\ \citenamefont {Tun}}]{Gaulin1992}%
  \BibitemOpen
  \bibfield  {author} {\bibinfo {author} {\bibfnamefont {B.~D.}\ \bibnamefont
  {Gaulin}}, \bibinfo {author} {\bibfnamefont {J.~N.}\ \bibnamefont {Reimers}},
  \bibinfo {author} {\bibfnamefont {T.~E.}\ \bibnamefont {Mason}}, \bibinfo
  {author} {\bibfnamefont {J.~E.}\ \bibnamefont {Greedan}}, \ and\ \bibinfo
  {author} {\bibfnamefont {Z.}~\bibnamefont {Tun}},\ }\href@noop {} {\bibfield
  {journal} {\bibinfo  {journal} {Phys. Rev. Lett.}\ }\textbf {\bibinfo
  {volume} {69}},\ \bibinfo {pages} {3244} (\bibinfo {year}
  {1992})}\BibitemShut {NoStop}%
\bibitem [{\citenamefont {Zhou}\ \emph {et~al.}(2010)\citenamefont {Zhou},
  \citenamefont {Wiebe}, \citenamefont {Janik}, \citenamefont {Vogt},
  \citenamefont {Harter}, \citenamefont {Dalal},\ and\ \citenamefont
  {Gardner}}]{ZHOU2010}%
  \BibitemOpen
  \bibfield  {author} {\bibinfo {author} {\bibfnamefont {H.}~\bibnamefont
  {Zhou}}, \bibinfo {author} {\bibfnamefont {C.}~\bibnamefont {Wiebe}},
  \bibinfo {author} {\bibfnamefont {J.}~\bibnamefont {Janik}}, \bibinfo
  {author} {\bibfnamefont {B.}~\bibnamefont {Vogt}}, \bibinfo {author}
  {\bibfnamefont {A.}~\bibnamefont {Harter}}, \bibinfo {author} {\bibfnamefont
  {N.}~\bibnamefont {Dalal}}, \ and\ \bibinfo {author} {\bibfnamefont
  {J.}~\bibnamefont {Gardner}},\ }\href@noop {} {\bibfield  {journal} {\bibinfo
   {journal} {Journal of Solid State Chemistry}\ }\textbf {\bibinfo {volume}
  {183}},\ \bibinfo {pages} {890 } (\bibinfo {year} {2010})}\BibitemShut
  {NoStop}%
\bibitem [{\citenamefont {Shimakawa}\ \emph {et~al.}(1999)\citenamefont
  {Shimakawa}, \citenamefont {Kubo}, \citenamefont {Hamada}, \citenamefont
  {Jorgensen}, \citenamefont {Hu}, \citenamefont {Short}, \citenamefont
  {Nohara},\ and\ \citenamefont {Takagi}}]{Shimakawa1999}%
  \BibitemOpen
  \bibfield  {author} {\bibinfo {author} {\bibfnamefont {Y.}~\bibnamefont
  {Shimakawa}}, \bibinfo {author} {\bibfnamefont {Y.}~\bibnamefont {Kubo}},
  \bibinfo {author} {\bibfnamefont {N.}~\bibnamefont {Hamada}}, \bibinfo
  {author} {\bibfnamefont {J.~D.}\ \bibnamefont {Jorgensen}}, \bibinfo {author}
  {\bibfnamefont {Z.}~\bibnamefont {Hu}}, \bibinfo {author} {\bibfnamefont
  {S.}~\bibnamefont {Short}}, \bibinfo {author} {\bibfnamefont
  {M.}~\bibnamefont {Nohara}}, \ and\ \bibinfo {author} {\bibfnamefont
  {H.}~\bibnamefont {Takagi}},\ }\href@noop {} {\bibfield  {journal} {\bibinfo
  {journal} {Phys. Rev. B}\ }\textbf {\bibinfo {volume} {59}},\ \bibinfo
  {pages} {1249} (\bibinfo {year} {1999})}\BibitemShut {NoStop}%
\bibitem [{\citenamefont {Subramanian}\ \emph {et~al.}(1988)\citenamefont
  {Subramanian}, \citenamefont {Torardi}, \citenamefont {Johnson},
  \citenamefont {Pannetier},\ and\ \citenamefont {Sleight}}]{SUBRAMANIAN1988}%
  \BibitemOpen
  \bibfield  {author} {\bibinfo {author} {\bibfnamefont {M.}~\bibnamefont
  {Subramanian}}, \bibinfo {author} {\bibfnamefont {C.}~\bibnamefont
  {Torardi}}, \bibinfo {author} {\bibfnamefont {D.}~\bibnamefont {Johnson}},
  \bibinfo {author} {\bibfnamefont {J.}~\bibnamefont {Pannetier}}, \ and\
  \bibinfo {author} {\bibfnamefont {A.}~\bibnamefont {Sleight}},\ }\href@noop
  {} {\bibfield  {journal} {\bibinfo  {journal} {Journal of Solid State
  Chemistry}\ }\textbf {\bibinfo {volume} {72}},\ \bibinfo {pages} {24 }
  (\bibinfo {year} {1988})}\BibitemShut {NoStop}%
\bibitem [{\citenamefont {Reimers}\ \emph {et~al.}(1988)\citenamefont
  {Reimers}, \citenamefont {Greedan},\ and\ \citenamefont
  {Sato}}]{REIMERS1988}%
  \BibitemOpen
  \bibfield  {author} {\bibinfo {author} {\bibfnamefont {J.}~\bibnamefont
  {Reimers}}, \bibinfo {author} {\bibfnamefont {J.}~\bibnamefont {Greedan}}, \
  and\ \bibinfo {author} {\bibfnamefont {M.}~\bibnamefont {Sato}},\ }\href@noop
  {} {\bibfield  {journal} {\bibinfo  {journal} {Journal of Solid State
  Chemistry}\ }\textbf {\bibinfo {volume} {72}},\ \bibinfo {pages} {390 }
  (\bibinfo {year} {1988})}\BibitemShut {NoStop}%
\bibitem [{\citenamefont {FLEURgroup}()}]{fleur}%
  \BibitemOpen
  \bibfield  {author} {\bibinfo {author} {\bibnamefont {FLEURgroup}},\ }\href
  {http://www.flapw.de/} {\enquote {\bibinfo {title} {http://www.flapw.de/},}\
  }\BibitemShut {NoStop}%
\bibitem [{\citenamefont {Giannozzi}\ \emph {et~al.}(2009)\citenamefont
  {Giannozzi}, \citenamefont {Baroni}, \citenamefont {Bonini}, \citenamefont
  {Calandra}, \citenamefont {Car}, \citenamefont {Cavazzoni}, \citenamefont
  {Ceresoli}, \citenamefont {Chiarotti}, \citenamefont {Cococcioni},
  \citenamefont {Dabo}, \citenamefont {Corso}, \citenamefont {de~Gironcoli},
  \citenamefont {Fabris}, \citenamefont {Fratesi}, \citenamefont {Gebauer},
  \citenamefont {Gerstmann}, \citenamefont {Gougoussis}, \citenamefont
  {Kokalj}, \citenamefont {Lazzeri}, \citenamefont {Martin-Samos},
  \citenamefont {Marzari}, \citenamefont {Mauri}, \citenamefont {Mazzarello},
  \citenamefont {Paolini}, \citenamefont {Pasquarello}, \citenamefont
  {Paulatto}, \citenamefont {Sbraccia}, \citenamefont {Scandolo}, \citenamefont
  {Sclauzero}, \citenamefont {Seitsonen}, \citenamefont {Smogunov},
  \citenamefont {Umari},\ and\ \citenamefont {Wentzcovitch}}]{Giannozzi_2009}%
  \BibitemOpen
  \bibfield  {author} {\bibinfo {author} {\bibfnamefont {P.}~\bibnamefont
  {Giannozzi}}, \bibinfo {author} {\bibfnamefont {S.}~\bibnamefont {Baroni}},
  \bibinfo {author} {\bibfnamefont {N.}~\bibnamefont {Bonini}}, \bibinfo
  {author} {\bibfnamefont {M.}~\bibnamefont {Calandra}}, \bibinfo {author}
  {\bibfnamefont {R.}~\bibnamefont {Car}}, \bibinfo {author} {\bibfnamefont
  {C.}~\bibnamefont {Cavazzoni}}, \bibinfo {author} {\bibfnamefont
  {D.}~\bibnamefont {Ceresoli}}, \bibinfo {author} {\bibfnamefont {G.~L.}\
  \bibnamefont {Chiarotti}}, \bibinfo {author} {\bibfnamefont {M.}~\bibnamefont
  {Cococcioni}}, \bibinfo {author} {\bibfnamefont {I.}~\bibnamefont {Dabo}},
  \bibinfo {author} {\bibfnamefont {A.~D.}\ \bibnamefont {Corso}}, \bibinfo
  {author} {\bibfnamefont {S.}~\bibnamefont {de~Gironcoli}}, \bibinfo {author}
  {\bibfnamefont {S.}~\bibnamefont {Fabris}}, \bibinfo {author} {\bibfnamefont
  {G.}~\bibnamefont {Fratesi}}, \bibinfo {author} {\bibfnamefont
  {R.}~\bibnamefont {Gebauer}}, \bibinfo {author} {\bibfnamefont
  {U.}~\bibnamefont {Gerstmann}}, \bibinfo {author} {\bibfnamefont
  {C.}~\bibnamefont {Gougoussis}}, \bibinfo {author} {\bibfnamefont
  {A.}~\bibnamefont {Kokalj}}, \bibinfo {author} {\bibfnamefont
  {M.}~\bibnamefont {Lazzeri}}, \bibinfo {author} {\bibfnamefont
  {L.}~\bibnamefont {Martin-Samos}}, \bibinfo {author} {\bibfnamefont
  {N.}~\bibnamefont {Marzari}}, \bibinfo {author} {\bibfnamefont
  {F.}~\bibnamefont {Mauri}}, \bibinfo {author} {\bibfnamefont
  {R.}~\bibnamefont {Mazzarello}}, \bibinfo {author} {\bibfnamefont
  {S.}~\bibnamefont {Paolini}}, \bibinfo {author} {\bibfnamefont
  {A.}~\bibnamefont {Pasquarello}}, \bibinfo {author} {\bibfnamefont
  {L.}~\bibnamefont {Paulatto}}, \bibinfo {author} {\bibfnamefont
  {C.}~\bibnamefont {Sbraccia}}, \bibinfo {author} {\bibfnamefont
  {S.}~\bibnamefont {Scandolo}}, \bibinfo {author} {\bibfnamefont
  {G.}~\bibnamefont {Sclauzero}}, \bibinfo {author} {\bibfnamefont {A.~P.}\
  \bibnamefont {Seitsonen}}, \bibinfo {author} {\bibfnamefont {A.}~\bibnamefont
  {Smogunov}}, \bibinfo {author} {\bibfnamefont {P.}~\bibnamefont {Umari}}, \
  and\ \bibinfo {author} {\bibfnamefont {R.~M.}\ \bibnamefont {Wentzcovitch}},\
  }\href@noop {} {\bibfield  {journal} {\bibinfo  {journal} {Journal of
  Physics: Condensed Matter}\ }\textbf {\bibinfo {volume} {21}},\ \bibinfo
  {pages} {395502} (\bibinfo {year} {2009})}\BibitemShut {NoStop}%
\bibitem [{\citenamefont {Garrity}\ \emph {et~al.}(2014)\citenamefont
  {Garrity}, \citenamefont {Bennett}, \citenamefont {Rabe},\ and\ \citenamefont
  {Vanderbilt}}]{Vanderbilt2014}%
  \BibitemOpen
  \bibfield  {author} {\bibinfo {author} {\bibfnamefont {K.~F.}\ \bibnamefont
  {Garrity}}, \bibinfo {author} {\bibfnamefont {J.~W.}\ \bibnamefont
  {Bennett}}, \bibinfo {author} {\bibfnamefont {K.~M.}\ \bibnamefont {Rabe}}, \
  and\ \bibinfo {author} {\bibfnamefont {D.}~\bibnamefont {Vanderbilt}},\
  }\href@noop {} {\bibfield  {journal} {\bibinfo  {journal} {Computational
  Materials Science}\ }\textbf {\bibinfo {volume} {81}},\ \bibinfo {pages} {446
  } (\bibinfo {year} {2014})}\BibitemShut {NoStop}%
\bibitem [{\citenamefont {Perdew}\ \emph {et~al.}(1996)\citenamefont {Perdew},
  \citenamefont {Burke},\ and\ \citenamefont {Ernzerhof}}]{Perdew1996}%
  \BibitemOpen
  \bibfield  {author} {\bibinfo {author} {\bibfnamefont {J.~P.}\ \bibnamefont
  {Perdew}}, \bibinfo {author} {\bibfnamefont {K.}~\bibnamefont {Burke}}, \
  and\ \bibinfo {author} {\bibfnamefont {M.}~\bibnamefont {Ernzerhof}},\
  }\href@noop {} {\bibfield  {journal} {\bibinfo  {journal} {Phys. Rev. Lett.}\
  }\textbf {\bibinfo {volume} {77}},\ \bibinfo {pages} {3865} (\bibinfo {year}
  {1996})}\BibitemShut {NoStop}%
\bibitem [{\citenamefont {de-la Roza}\ \emph {et~al.}(2014)\citenamefont {de-la
  Roza}, \citenamefont {Johnson},\ and\ \citenamefont {Luaña}}]{critic2}%
  \BibitemOpen
  \bibfield  {author} {\bibinfo {author} {\bibfnamefont {A.~O.}\ \bibnamefont
  {de-la Roza}}, \bibinfo {author} {\bibfnamefont {E.~R.}\ \bibnamefont
  {Johnson}}, \ and\ \bibinfo {author} {\bibfnamefont {V.}~\bibnamefont
  {Luaña}},\ }\href {\doibase http://dx.doi.org/10.1016/j.cpc.2013.10.026}
  {\bibfield  {journal} {\bibinfo  {journal} {Computer Physics Communications}\
  }\textbf {\bibinfo {volume} {185}},\ \bibinfo {pages} {1007 } (\bibinfo
  {year} {2014})}\BibitemShut {NoStop}%
\bibitem [{\citenamefont {de-la Roza}\ \emph {et~al.}(2009)\citenamefont {de-la
  Roza}, \citenamefont {Blanco}, \citenamefont {Pendás},\ and\ \citenamefont
  {Luaña}}]{critic}%
  \BibitemOpen
  \bibfield  {author} {\bibinfo {author} {\bibfnamefont {A.~O.}\ \bibnamefont
  {de-la Roza}}, \bibinfo {author} {\bibfnamefont {M.}~\bibnamefont {Blanco}},
  \bibinfo {author} {\bibfnamefont {A.~M.}\ \bibnamefont {Pendás}}, \ and\
  \bibinfo {author} {\bibfnamefont {V.}~\bibnamefont {Luaña}},\ }\href
  {\doibase http://dx.doi.org/10.1016/j.cpc.2008.07.018} {\bibfield  {journal}
  {\bibinfo  {journal} {Computer Physics Communications}\ }\textbf {\bibinfo
  {volume} {180}},\ \bibinfo {pages} {157 } (\bibinfo {year}
  {2009})}\BibitemShut {NoStop}%
\bibitem [{\citenamefont {Anisimov}\ \emph {et~al.}(1991)\citenamefont
  {Anisimov}, \citenamefont {Zaanen},\ and\ \citenamefont
  {Andersen}}]{Anisimov1991}%
  \BibitemOpen
  \bibfield  {author} {\bibinfo {author} {\bibfnamefont {V.~I.}\ \bibnamefont
  {Anisimov}}, \bibinfo {author} {\bibfnamefont {J.}~\bibnamefont {Zaanen}}, \
  and\ \bibinfo {author} {\bibfnamefont {O.~K.}\ \bibnamefont {Andersen}},\
  }\href@noop {} {\bibfield  {journal} {\bibinfo  {journal} {Phys. Rev. B}\
  }\textbf {\bibinfo {volume} {44}},\ \bibinfo {pages} {943} (\bibinfo {year}
  {1991})}\BibitemShut {NoStop}%
\bibitem [{\citenamefont {Vaugier}\ \emph {et~al.}(2012)\citenamefont
  {Vaugier}, \citenamefont {Jiang},\ and\ \citenamefont {Biermann}}]{Vaugier}%
  \BibitemOpen
  \bibfield  {author} {\bibinfo {author} {\bibfnamefont {L.}~\bibnamefont
  {Vaugier}}, \bibinfo {author} {\bibfnamefont {H.}~\bibnamefont {Jiang}}, \
  and\ \bibinfo {author} {\bibfnamefont {S.}~\bibnamefont {Biermann}},\ }\href
  {\doibase 10.1103/PhysRevB.86.165105} {\bibfield  {journal} {\bibinfo
  {journal} {Phys. Rev. B}\ }\textbf {\bibinfo {volume} {86}},\ \bibinfo
  {pages} {165105} (\bibinfo {year} {2012})}\BibitemShut {NoStop}%
\bibitem [{\citenamefont {Cococcioni}\ and\ \citenamefont
  {de~Gironcoli}(2005)}]{Cococcioni2005}%
  \BibitemOpen
  \bibfield  {author} {\bibinfo {author} {\bibfnamefont {M.}~\bibnamefont
  {Cococcioni}}\ and\ \bibinfo {author} {\bibfnamefont {S.}~\bibnamefont
  {de~Gironcoli}},\ }\href@noop {} {\bibfield  {journal} {\bibinfo  {journal}
  {Phys. Rev. B}\ }\textbf {\bibinfo {volume} {71}},\ \bibinfo {pages} {035105}
  (\bibinfo {year} {2005})}\BibitemShut {NoStop}%
\bibitem [{\citenamefont {Kulik}\ \emph {et~al.}(2006)\citenamefont {Kulik},
  \citenamefont {Cococcioni}, \citenamefont {Scherlis},\ and\ \citenamefont
  {Marzari}}]{Kulik2006}%
  \BibitemOpen
  \bibfield  {author} {\bibinfo {author} {\bibfnamefont {H.~J.}\ \bibnamefont
  {Kulik}}, \bibinfo {author} {\bibfnamefont {M.}~\bibnamefont {Cococcioni}},
  \bibinfo {author} {\bibfnamefont {D.~A.}\ \bibnamefont {Scherlis}}, \ and\
  \bibinfo {author} {\bibfnamefont {N.}~\bibnamefont {Marzari}},\ }\href@noop
  {} {\bibfield  {journal} {\bibinfo  {journal} {Phys. Rev. Lett.}\ }\textbf
  {\bibinfo {volume} {97}},\ \bibinfo {pages} {103001} (\bibinfo {year}
  {2006})}\BibitemShut {NoStop}%
\bibitem [{\citenamefont {Sadeghi}\ \emph {et~al.}(2015)\citenamefont
  {Sadeghi}, \citenamefont {Alaei}, \citenamefont {Shahbazi},\ and\
  \citenamefont {Gingras}}]{sadeghi2015}%
  \BibitemOpen
  \bibfield  {author} {\bibinfo {author} {\bibfnamefont {A.}~\bibnamefont
  {Sadeghi}}, \bibinfo {author} {\bibfnamefont {M.}~\bibnamefont {Alaei}},
  \bibinfo {author} {\bibfnamefont {F.}~\bibnamefont {Shahbazi}}, \ and\
  \bibinfo {author} {\bibfnamefont {M.~J.~P.}\ \bibnamefont {Gingras}},\
  }\href@noop {} {\bibfield  {journal} {\bibinfo  {journal} {Phys. Rev. B}\
  }\textbf {\bibinfo {volume} {91}},\ \bibinfo {pages} {140407(R)} (\bibinfo
  {year} {2015})}\BibitemShut {NoStop}%
\bibitem [{\citenamefont {Moriya}(1960)}]{Moriya1960}%
  \BibitemOpen
  \bibfield  {author} {\bibinfo {author} {\bibfnamefont {T.}~\bibnamefont
  {Moriya}},\ }\href@noop {} {\bibfield  {journal} {\bibinfo  {journal} {Phys.
  Rev.}\ }\textbf {\bibinfo {volume} {120}},\ \bibinfo {pages} {91} (\bibinfo
  {year} {1960})}\BibitemShut {NoStop}%
\bibitem [{\citenamefont {Hukushima}\ and\ \citenamefont
  {Nemoto}(1996)}]{Hukushima1996}%
  \BibitemOpen
  \bibfield  {author} {\bibinfo {author} {\bibfnamefont {K.}~\bibnamefont
  {Hukushima}}\ and\ \bibinfo {author} {\bibfnamefont {K.}~\bibnamefont
  {Nemoto}},\ }\href@noop {} {\bibfield  {journal} {\bibinfo  {journal}
  {Journal of the Physical Society of Japan}\ }\textbf {\bibinfo {volume}
  {65}},\ \bibinfo {pages} {1604} (\bibinfo {year} {1996})}\BibitemShut
  {NoStop}%
\bibitem [{\citenamefont {Rezaei}\ \emph {et~al.}(2019)\citenamefont {Rezaei},
  \citenamefont {Alaei},\ and\ \citenamefont {Akbarzadeh}}]{rezaei2019espins}%
  \BibitemOpen
  \bibfield  {author} {\bibinfo {author} {\bibfnamefont {N.}~\bibnamefont
  {Rezaei}}, \bibinfo {author} {\bibfnamefont {M.}~\bibnamefont {Alaei}}, \
  and\ \bibinfo {author} {\bibfnamefont {H.}~\bibnamefont {Akbarzadeh}},\
  }\href@noop {} {\  (\bibinfo {year} {2019})},\ \Eprint
  {http://arxiv.org/abs/1912.00793} {arXiv:1912.00793 [physics.comp-ph]}
  \BibitemShut {NoStop}%
\bibitem [{\citenamefont {Elhajal}\ \emph {et~al.}(2005)\citenamefont
  {Elhajal}, \citenamefont {Canals}, \citenamefont {Sunyer},\ and\
  \citenamefont {Lacroix}}]{Elhajal2005}%
  \BibitemOpen
  \bibfield  {author} {\bibinfo {author} {\bibfnamefont {M.}~\bibnamefont
  {Elhajal}}, \bibinfo {author} {\bibfnamefont {B.}~\bibnamefont {Canals}},
  \bibinfo {author} {\bibfnamefont {R.}~\bibnamefont {Sunyer}}, \ and\ \bibinfo
  {author} {\bibfnamefont {C.}~\bibnamefont {Lacroix}},\ }\href@noop {}
  {\bibfield  {journal} {\bibinfo  {journal} {Phys. Rev. B}\ }\textbf {\bibinfo
  {volume} {71}},\ \bibinfo {pages} {094420} (\bibinfo {year}
  {2005})}\BibitemShut {NoStop}%
\bibitem [{\citenamefont {Donnerer}\ \emph {et~al.}(2016)\citenamefont
  {Donnerer}, \citenamefont {Rahn}, \citenamefont {Sala}, \citenamefont {Vale},
  \citenamefont {Pincini}, \citenamefont {Strempfer}, \citenamefont {Krisch},
  \citenamefont {Prabhakaran}, \citenamefont {Boothroyd},\ and\ \citenamefont
  {McMorrow}}]{Donnerer2016}%
  \BibitemOpen
  \bibfield  {author} {\bibinfo {author} {\bibfnamefont {C.}~\bibnamefont
  {Donnerer}}, \bibinfo {author} {\bibfnamefont {M.~C.}\ \bibnamefont {Rahn}},
  \bibinfo {author} {\bibfnamefont {M.~M.}\ \bibnamefont {Sala}}, \bibinfo
  {author} {\bibfnamefont {J.~G.}\ \bibnamefont {Vale}}, \bibinfo {author}
  {\bibfnamefont {D.}~\bibnamefont {Pincini}}, \bibinfo {author} {\bibfnamefont
  {J.}~\bibnamefont {Strempfer}}, \bibinfo {author} {\bibfnamefont
  {M.}~\bibnamefont {Krisch}}, \bibinfo {author} {\bibfnamefont
  {D.}~\bibnamefont {Prabhakaran}}, \bibinfo {author} {\bibfnamefont {A.~T.}\
  \bibnamefont {Boothroyd}}, \ and\ \bibinfo {author} {\bibfnamefont {D.~F.}\
  \bibnamefont {McMorrow}},\ }\href@noop {} {\bibfield  {journal} {\bibinfo
  {journal} {Phys. Rev. Lett.}\ }\textbf {\bibinfo {volume} {117}},\ \bibinfo
  {pages} {037201} (\bibinfo {year} {2016})}\BibitemShut {NoStop}%
\bibitem [{\citenamefont {Shinaoka}\ \emph {et~al.}(2013)\citenamefont
  {Shinaoka}, \citenamefont {Motome}, \citenamefont {Miyake},\ and\
  \citenamefont {Ishibashi}}]{Shinaoka2013}%
  \BibitemOpen
  \bibfield  {author} {\bibinfo {author} {\bibfnamefont {H.}~\bibnamefont
  {Shinaoka}}, \bibinfo {author} {\bibfnamefont {Y.}~\bibnamefont {Motome}},
  \bibinfo {author} {\bibfnamefont {T.}~\bibnamefont {Miyake}}, \ and\ \bibinfo
  {author} {\bibfnamefont {S.}~\bibnamefont {Ishibashi}},\ }\href {\doibase
  10.1103/PhysRevB.88.174422} {\bibfield  {journal} {\bibinfo  {journal} {Phys.
  Rev. B}\ }\textbf {\bibinfo {volume} {88}},\ \bibinfo {pages} {174422}
  (\bibinfo {year} {2013})}\BibitemShut {NoStop}%
\bibitem [{\citenamefont {Subramanian}\ \emph {et~al.}(1980)\citenamefont
  {Subramanian}, \citenamefont {Aravamudan},\ and\ \citenamefont
  {Rao}}]{Subramanian1980}%
  \BibitemOpen
  \bibfield  {author} {\bibinfo {author} {\bibfnamefont {M.}~\bibnamefont
  {Subramanian}}, \bibinfo {author} {\bibfnamefont {G.}~\bibnamefont
  {Aravamudan}}, \ and\ \bibinfo {author} {\bibfnamefont {G.~S.}\ \bibnamefont
  {Rao}},\ }\href {\doibase https://doi.org/10.1016/0025-5408(80)90094-X}
  {\bibfield  {journal} {\bibinfo  {journal} {Materials Research Bulletin}\
  }\textbf {\bibinfo {volume} {15}},\ \bibinfo {pages} {1401 } (\bibinfo {year}
  {1980})}\BibitemShut {NoStop}%
\bibitem [{\citenamefont {Meredig}\ \emph {et~al.}(2010)\citenamefont
  {Meredig}, \citenamefont {Thompson}, \citenamefont {Hansen}, \citenamefont
  {Wolverton},\ and\ \citenamefont {van~de Walle}}]{Meredig2010}%
  \BibitemOpen
  \bibfield  {author} {\bibinfo {author} {\bibfnamefont {B.}~\bibnamefont
  {Meredig}}, \bibinfo {author} {\bibfnamefont {A.}~\bibnamefont {Thompson}},
  \bibinfo {author} {\bibfnamefont {H.~A.}\ \bibnamefont {Hansen}}, \bibinfo
  {author} {\bibfnamefont {C.}~\bibnamefont {Wolverton}}, \ and\ \bibinfo
  {author} {\bibfnamefont {A.}~\bibnamefont {van~de Walle}},\ }\href@noop {}
  {\bibfield  {journal} {\bibinfo  {journal} {Phys. Rev. B}\ }\textbf {\bibinfo
  {volume} {82}},\ \bibinfo {pages} {195128} (\bibinfo {year}
  {2010})}\BibitemShut {NoStop}%
\bibitem [{\citenamefont {Deilynazar}\ \emph {et~al.}(2015)\citenamefont
  {Deilynazar}, \citenamefont {Khorasani}, \citenamefont {Alaei},\ and\
  \citenamefont {Hashemifar}}]{DEILYNAZAR2015}%
  \BibitemOpen
  \bibfield  {author} {\bibinfo {author} {\bibfnamefont {N.}~\bibnamefont
  {Deilynazar}}, \bibinfo {author} {\bibfnamefont {E.}~\bibnamefont
  {Khorasani}}, \bibinfo {author} {\bibfnamefont {M.}~\bibnamefont {Alaei}}, \
  and\ \bibinfo {author} {\bibfnamefont {S.~J.}\ \bibnamefont {Hashemifar}},\
  }\href {\doibase https://doi.org/10.1016/j.jmmm.2015.05.042} {\bibfield
  {journal} {\bibinfo  {journal} {Journal of Magnetism and Magnetic Materials}\
  }\textbf {\bibinfo {volume} {393}},\ \bibinfo {pages} {127 } (\bibinfo {year}
  {2015})}\BibitemShut {NoStop}%
\end{thebibliography}%
\end{document}